\documentclass{raa}
\usepackage{graphicx,times}
\usepackage{natbib}
\usepackage{amssymb,amsmath}
\bibpunct{(}{)}{;}{a}{}{,}

\usepackage{upgreek}

\def\kms  {km\,s$^{-1}$}

\def\deg  {\ifmmode {^\circ}\else {$^\circ$}\fi}
\def\porm {\ifmmode {\pm}\else {$\pm$}\fi}
\def\chisqpdf {\ifmmode {\chi^2_{\rm pdf}}\else {$\chi^2_{\rm pdf}$}\fi}
\def\chisq    {\ifmmode {\chi^2}\else {$\chi^2$}\fi}

\def\d    {\ifmmode {{\rlap{.}}^\circ}\else {${\rlap{.}}^\circ$}\fi}
\def\s    {\ifmmode {{\rlap{.}}^s}\else {${\rlap{.}}^s$}\fi}
\def\as   {\ifmmode {{\rlap{.}}^{''}}\else {${\rlap{.}}^{''}$}\fi}
\def\pa    {\ifmmode {\psi} \else {$\psi$}\fi}

\def\vlsr  {\ifmmode {v_{\rm LSR}}\else {$v_{\rm LSR}$}\fi}
\def\vlsrr {\ifmmode {v^r_{\rm LSR}}\else {$v^r_{\rm LSR}$}\fi}
\def\vhelio{\ifmmode {v_{Helio}}\else {$v_{Helio}$}\fi}

\def\ura   {\ifmmode {\mu_\alpha}\else {$\mu_\alpha$}\fi}
\def\udec  {\ifmmode {\mu_\delta}\else {$\mu_\delta$}\fi}

\def\ul    {\ifmmode {\mu_l}\else {$\mu_l$}\fi}
\def\ub    {\ifmmode {\mu_b}\else {$\mu_b$}\fi}

\def\uml   {\ifmmode {v_{gr}}\else {$v_{gr}$}\fi}
\def\umb   {\ifmmode {v_b}\else {$v_b$}\fi}
\def\vsrad {\ifmmode {v_{rad}}\else {$v_{rad}$}\fi}

\def\upl   {\ifmmode {v^p_{gr}}\else {$v^p_{gr}$}\fi}
\def\upb   {\ifmmode {v^p_b}\else {$v^p_b$}\fi}
\def\vprad {\ifmmode {v^p_{rad}}\else {$v^p_{rad}$}\fi}

\def\Vo    {\ifmmode {V^{Std}_\odot}\else {$V^{Std}_\odot$}\fi}
\def\Uo    {\ifmmode {U^{Std}_\odot}\else {$U^{Std}_\odot$}\fi}
\def\Wo    {\ifmmode {W^{Std}_\odot}\else {$W^{Std}_\odot$}\fi}
\def\VH    {\ifmmode {V^H_\odot}\else {$V^H_\odot$}\fi}
\def\UH    {\ifmmode {U^H_\odot}\else {$U^H_\odot$}\fi}
\def\WH    {\ifmmode {W^H_\odot}\else {$W^H_\odot$}\fi}
\def\V     {\ifmmode {V_\odot}\else {$V_\odot$}\fi}
\def\U     {\ifmmode {U_\odot}\else {$U_\odot$}\fi}
\def\W     {\ifmmode {W_\odot}\else {$W_\odot$}\fi}

\def\Vs    {\ifmmode {V_s}\else {$V_s$}\fi}
\def\Us    {\ifmmode {U_s}\else {$U_s$}\fi}
\def\Ws    {\ifmmode {W_s}\else {$W_s$}\fi}

\def\Vsbar {\ifmmode {\overline{V_s}}\else {$\overline{V_s}$}\fi}
\def\Usbar {\ifmmode {\overline{U_s}}\else {$\overline{U_s}$}\fi}
\def\Wsbar {\ifmmode {\overline{W_s}}\else {$\overline{W_s}$}\fi}

\def\pars  {\ifmmode{\pi_s}\else{$\pi_s$}\fi}

\def\Ts    {\ifmmode{\Theta_s}\else{$\Theta_s$}\fi}
\def\Tdot  {\ifmmode{d\Theta\over dR}\else{$d\Theta\over dR$}\fi}

\def\Rp    {\ifmmode{R_p}\else{$R_p$}\fi}

\def\To    {\ifmmode{\Theta_0}\else{$\Theta_0$}\fi}
\def\Ro    {\ifmmode{R_0}\else{$R_0$}\fi}
\def\Vlsr {\ifmmode {V_{\rm LSR}} \else {$V_{\rm LSR}$} \fi}

 \renewcommand{\thepage}{146--\arabic{page}}
\begin{document}

   \title{The spiral structure of the Milky Way}

\volnopage{{\bf 2018} Vol.~{\bf 18} No.~{\bf 12}, ~146(20pp)~
   {\small  doi: 10.1088/1674--4527/18/12/146}}      %%preserved for Editor. DOn't remove!
   \setcounter{page}{1}

   \author{Ye Xu\inst{1}
   \and Li-Gang Hou\inst{2,3}
   \and Yuan-Wei Wu\inst{4}
  }

   \institute{Purple Mountain Observatory, Chinese Academy of
Sciences, Nanjing 210008, China; {\it xuye@pmo.ac.cn}\\
\and
CAS Key Laboratory of FAST, National Astronomical Observatories, Chinese Academy of Sciences, Beijing 100101, China\\
       \and
       National Astronomical Observatories, Chinese Academy of Sciences, Beijing 100101, China\\
       \and
       National Time Service Center, Key Laboratory of Precise
Positioning and Timing Technology, Chinese Academy of Sciences, Xi'an 710600,
China\\
    \vs \no
   {\small Received 2018 September 3; accepted 2018 October 9}
}

\abstract{  The morphology and kinematics of the spiral structure of
the Milky Way {are} long-standing problem{s} in
astrophysics. In this review
  we firstly summarize various methods with different tracers used to
  solve this puzzle. The astrometry of Galactic sources is gradually
  alleviating this difficult situation
  caused mainly by large distance uncertainties, as
  we can currently obtain accurate parallaxes (a few {$\upmu$as}) and
  proper motions ($\approx$1{\,}\kms) by using Very Long Baseline
  Interferometry (VLBI). On the other hand, {the }{{\it Gaia}} mission is providing
  the largest, uniform sample of parallaxes for O-type stars in the
  entire Milky Way. Based upon the VLBI maser and {{\it Gaia}} O-star parallax
  measurements, nearby spiral structures{ of} the Perseus, Local, Sagittarius
  and Scutum {A}rms are determined in unprecedented detail. Meanwhile,
  we estimate fundamental Galactic parameters{ of} the
  distance to the Galactic center, \Ro, to be $8.35\pm0.18${\,}kpc,
  a{nd}
  circular rotation speed at the Sun, \To, to be 240$\pm$10{\,}\kms.
  We found kinematic differences between O stars and
interstellar masers: the O stars, on average, rotate
      faster, $>$8{\,}\kms\ than maser-traced high-mass star forming regions.
}

   \authorrunning{{\it Y. Xu et al}.: ~The Spiral Structure of the Milky Way  }            %author_head in even pages
   \titlerunning{{\it Y. Xu et al}.: ~The Spiral Structure of the Milky Way}  % title_head in odd pages
   \maketitle
%________________________________________________ sections below
%
\section{Introduction}           %% first-level sections will be auto-capitalized
\label{sect:intro}

The Milky Way was proposed to be a spiral galaxy soon after the
discovery of spiral structures in M51 \citep{Alexander:52} more than
one and a half centuries ago. However, the Galactic spiral structure
is extremely difficult to depict because the Solar System is deeply
embedded in the Galactic disk. Galactic rotation was
revealed by Oort in the 1920s \citep{1927BAN.....3..275O}, and a
major breakthrough towards understanding the Galactic spiral
structure happened in the 1950s -- Morgan and his colleagues found
three spiral arm segments in the solar neighborhood using photometry
\citep{Mor:52,Mor:53}. Unfortunately, in the optical band,
interstellar dust along the line of sight prevents us from
determining{ the} large scale Galactic spiral pattern beyond a
few kpc {from} the Sun. Alternatively, observations in
radio bands, e.g., HI and CO molecul{ar} lines, which are
free from being affected by dust extinction, offer new
opportunities to investigate the Galactic spiral structure. However,
the kinematic distances derived from rotation curves have large
errors, imposing large uncertainties on the identification of spiral
arms.

More than 100 models have been proposed to explain the Galactic
spiral pattern, but most of them employed kinematic
distances. The uncertainties mainly come from three causes: (1)
difficulties in determining an accurate rotation curve, (2)
kinematic distance ambiguities\footnote{For a source in the inner
Galaxy whose distance
  to the GC is less than the distance between the Sun and the GC,
  $R_0$, there exist two possible distances corresponding to one
  observed velocity with respect to the Local Standard of Rest,
  $V_{\rm LSR}$.}, and (3) deviations from non-circular rotation (e.g.,
streaming motions). These factors yield uncertainties comparable to
the gaps between arms. For instance, for the molecular cloud
G9.62+0.20, its far and near kinematic distances are approximately
15 and 0.5{\,}kpc, respectively, but its true distance is
about 5.7{\,}kpc \citep{sanna09}. Therefore, it is
hard to determine precise locations of molecular clouds and to
construct the morphology of Galactic spiral arms. Up to now, there
is no general consensus on the number of arms, their locations,
orientations or properties.

Recently, substantial progress in our knowledge of the spatial and
kinematic properties of Galactic structure has been
achieved. For example, \citet{xrzm06} and \citet{Honma:07}
demonstrated that Very Long Baseline Interferometry (VLBI) can
obtain trigonometric parallax accuracies down to a few $\upmu$as,
allowing precise distance measurements towards masers throughout the
Galaxy, which was recognized as a milestone in this field
\citep{binney06,cas+12}. Large portions of spiral arms have now been
accurately defined in the northern hemisphere
\citep[e.g.,][]{Reid:14,xrd+16}, and in addition, the distance to
the Galactic Center (GC) and the Galactic rotation speed at the Sun
have been well determined \citep{Honma:12,Reid:14}.

On the other hand, the {{\it Gaia}} satellite,
launched in 2013, is collecting the most precise astrometric
measurements for billions of stars in the Milky Way, and the
{{\it Gaia}} mission recently released its second data
set ({Data Release 2, }DR2), containing more than one billion
stars that have parallaxes and proper motions measured by
\cite{DR1,DR2}. The parallax uncertainty in {{\it
Gaia}} DR2 is typically 30{\,}$\upmu$as. With a large
number of parallax-measured OB stars, the spiral pattern within
3\,kpc from the Sun could be revealed clearly for the first time
\citep{Xu+18}.

In this review, we present the results from multiple tracers
proposed over the past half century, including ionized hydrogen,
neutral atomic hydrogen, molecular gas, young open clusters
and particularly the results from maser trigonometric parallax.
Additionally, we describe our latest research results about the
stretch of spiral arms and their space motions{,} and Galactic
fundamental parameters based on maser and O-type star parallax and
proper motion measurements.

\section{An overview of various tracers}

There are roughly two kinds of spiral arms, one of which is
associated with young objects, such as OB stars and young stellar
associations. Such spiral arms are birthplaces of star{s}, and
consequently, giant molecular clouds (GMCs), young open
clusters and photodissociation regions are{ the} best
tracers of this type of arm. However, the other kind is
mainly traced by more evolved stars, which have moved out of their
birthplaces and form another kind of spiral arm. In addition, atomic
gas is characterized by 21{\,}cm HI emissions{
which}, due to its wide range, trace{s} spiral arms on a
larger scale. In external galaxies{,} the spiral arms
traced by HI are similar to CO molecular gas in general, but with
detailed differences \citep{wes98}. However, most researchers
prefer combining all the tracers to just outline a single
spiral arm pattern.

\subsection{OB Stars and Their HII Regions}
\label{sect:Ob}

The first spiral structure {was} found in M51, a nearby
galaxy, based on observations of high-mass stars and bright HII
regions \citep{1850RSPT..140..499R}. In the Milky Way, OB stars and
young stellar associations are also primary tracers of spiral arms,
especially their associated HII regions, which are bright in radio
wavelength{s} and almost immune to interstellar dust
extinction, and they can be widely detected throughout the entire
Galactic plane \citep[e.g., as far as more than 20\,kpc away from
the
  observer,][]{abbr12}.

The global spiral arms depicted by massive young stellar objects
(MYSOs){ }\citep[e.g.,][]{gg76,rus03,hh14} provide {a
}starting point for some well-known models of the Milky Way, e.g.,
the electron-density models \citep[][]{ne2001,ymw17}, the model of
dust distribution \citep[][]{ds01} and the large-scale
magnetic field structure model throughout the Galactic disk
\citep[e.g.,][]{ hmvd18}. The distances of spiral tracers are key
parameters to map the Galaxy{'s} spiral arms. The most reliable
and direct method of determining the distances of MYSOs is to
measure the trigonometric parallax of their associated
methanol/water masers \citep[e.g.,][]{xrzm06,hbm+06}. The
spectrophotometry of high-mass stars in HII regions, which
{is} based on interstellar extinction laws, is also a
good method and has determined stellar distances for about 400 HII
regions \citep[e.g.,][]{rus03,fb15}. For large samples \citep[more
than 1200,
  e.g., see][]{hh14} of Galactic HII regions and masers in
high-mass star forming regions{ (HMSFRs)}, only kinematic
distances were estimated from their $V_{\rm LSR}$ by using a mean
Galaxy rotation curve.

\begin{figure*}
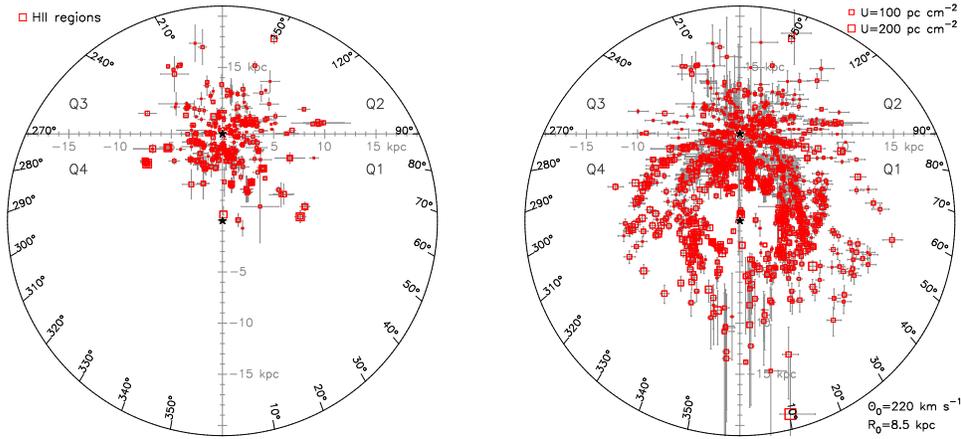
  %%fig1
\centering
\includegraphics[width=0.4\textwidth]{1.eps}\hspace{1cm}
\includegraphics[width=0.4\textwidth]{2.eps}
\vspace{0.5cm}

\caption{\baselineskip 3.8mm {\it Left:} Distribution of HII regions
({\it red}) with known
  spectrophotometric distances. {\it Right:} Distribution of Galactic
  HII regions ({\it red}) with spectroscopic distances or kinematic
  distances. The symbol size is proportional to the excitation
  parameters. The IAU standard $R_0=8.5${\,}kpc and
  $\Theta_0=220${\,}km{\,}s$^{-1}${,} and the standard solar motions together
  with a flat rotation curve are adopted in deriving the kinematic
  distances. Two {\it black stars} indicate the location{s} of the Sun
  ($x=$~0.0{\,}kpc, $y=$~8.5{\,}kpc) and the GC ($x=$~0.0{\,}kpc,
  $y=$~0.0{\,}kpc). Q1 to Q4 indicate the four Galactic
  quadrants. Position uncertainties are indicated by error bars
  ({\it gray}). Galactic longitudes in degrees are also marked in the
  plots. The HII region data are taken from \citet{hh14}.}
\label{hii}
\end{figure*}

We briefly review the time line of Galactic spiral arm studies.
Using O and early B stars, Morgan and his collaborators
first outlined parts of nearby spiral arms, three short
spiral arm segments, with spectroscopic parallaxes
\citep[e.g.,][]{Mor:52,Mor:53}. Based on the distributions of OB
stars and optical/radio HII regions, \citet{bhm70} mapped the Carina
spiral feature in Galactic longitude from 285$^\circ$ to
295$^\circ$. Using a sample of about 160 HII regions with
spectrophotometric or kinematic distances, \citet{cg75} identified
four spiral arm segments. \citet{gg76} determined the positions of
100 clusters of HII regions by spectrophotometric or kinematic
distances of 360 exciting stars. These HII regions are proposed to
reside in part of four spiral arms, i.e., the Perseus Arm, the
Sagittarius-Carina Arm, the Scutum-Crux-Centaurus Arm and the Norma Arm. \citet{fb84} made a similar map but using more
than 100{ }HII regions with spectrophotometric distances
(their fig.~2). \citet{ave85} obtained four clear spiral arm
segments within about 6{\,}kpc of the Sun (their fig.~3)
with spectrophotometric distances known for 255 HII regions, but the
data {were} not released. \citet{rus03} updated the
stellar distances for 204 star forming complexes. \citet{fb15}
determined the spectrophotometric distances to 103 HII regions in
the second and third Galactic quadrants. The distance accuracy of
the spectrophotometric method is not as good as that of
trigonometric parallax, but the spectrophotometric method still
provide{s} relatively accurate distances with uncertainties of
about 20\% \citep[e.g.,][]{rus03}, and has been used to measure the
mean Galaxy rotation curves \citep[e.g.,][]{bb93,rus03}. Due to absorption by dust, optical methods are limited to
nearby spiral structures and {are }ineffective at distances
greater than a few kpc, as
shown in Figure~\ref{hii} %1111111111
 left. About 400 HII regions in
total{,} within about 6\,kpc of the Sun, trace part of the
Perseus Arm, the Local Arm, the Sagittarius-Carina Arm and
the Scutum-Crux-Centaurus Arm.

The paradigmatic map of Galaxy spiral arms was given by
\citet[][]{gg76} who first proposed that the Milky Way probably has
four major spiral arms. It is noted that in their model, the Local
Arm was a spur or a branch, not a major arm. \citet{dwbw80} and
\citet{ch87} extended the four-arm model by observing 171 and 316
HII regions in the northern and southern sky, respectively. To
update the global maps of spiral arms, \citet{rus03} cataloged 481
star forming complexes and determined their spectrophotometric or
kinematic distances. The fitted model confirmed the four-segment
model of \citet{gg76}. Similar work was done by \citet{pdd04} with
550 HII regions, by \citet{hhs09} with 814 HII regions and
also by \citet{ufm+14} with about 1750 embedded young massive stars.
An up to date global picture of Gala{ctic} spiral arms
was given by \citet{hh14} with more than 1800 HII regions with known
trigonometric,
  spectrophotometric or kinematic distances. Based on the
distribution of known Galactic HII regions (Fig.~\ref{hii} right),
spiral arm segments are prominent in the first and fourth Galactic
quadrants, implying the existence of a coherent spiral pattern of
the Milky Way. Meanwhile, the HII region distribution is messy in
some Galaxy regions,{ and} the connections and continuity of arm
segments in different Galactic quadrants are still unclear.
Different models, e.g., three-arm and four-arm ones, are able to
connect most HII regions \citep[][]{hh14}. The classic four-arm
picture originally proposed by \citet{gg76} is unusually clean in
comparison with modern HII region maps \citep[e.g.,][]{fc10} and
seems not to be the unique solution. The Galaxy{'s} spiral
structure is far from a closed subject. To explicitly uncover the
entire picture, it is crucial to discover more weak and distant HII
regions \citep[e.g.,][]{aaj+15}, and reduce distance errors.

\subsection{Neutral Atomic Hydrogen}
Neutral atomic hydrogen (21 cm line) is ubiquitous in the Milky Way.
The well defined HI gas disk is suggested to extend to about
35{\,}kpc from the GC.  Structures at multiple scales are
present in the HI disk, from {a} small{
}scale, e.g., filaments, bubbles, shells{ and} spurs, to
{a} large{ }scale, such as warped,
flared features and also spiral arms. The HI gas can be mapped
throughout the entire Galaxy with the HI 21-cm line, providing a key
probe to study the structure and dynamics of the Milky Way
\citep[e.g., see][]{dl90,kk09}.

Soon after the discovery of{ the} Galactic HI 21-cm line
\citep[][]{ep51}, HI surveys were used to study the
large-scale spiral structure. \citet{ch52} found two separate long
features over a considerable range of Galactic longitudes in the
$l-v$ diagram, suggesting the possible existence of spiral arms.
Afterwards, early HI surveys were extended to a larger
portion of the Galactic disk, and the observed HI ($l$-$b$-$v$) data
were converted to neutral atomic gas distributions in the Galactic
plane \citep[e.g.,][]{vmo54,khc57,west57,okw58,bok59,wea70} using
velocity field models to derive the kinematic distances
\citep[e.g.,][]{sch56}, usually with an assumption of circular
rotation. Prominent features in their results are arm-like segments
extend{ing} from $R\sim3${\,}kpc to
$R>10${\,}kpc. Since then, studies on the
Galactic spiral structure {were} no longer confined to
the vicinity of the Sun by{ the} optical method
\citep[][]{Mor:52,Mor:53}, but extended to almost the entire
Galactic disk.

Along with the significant progresses, debates continue about maps of{ the} HI distribution, primarily about the inner
Galaxy regions \citep[][]{sim70}. Even with almost identical data,
derived HI maps of the inner Galaxy show discrepancies in
the number and position of spiral arms \citep[][]{ker69,wea70}.
Major causes are large uncertainties in the kinematic
distances of HI gas. Although Galactic HI primarily {has} circular rotation, random and
non-circular motions could be significant, as large as $\lesssim$
10{\,}km{\,}s$^{-1}$ or $<5\%$ of the
rotational velocity \citep[][]{lock02}. The velocity
crowding\footnote{The radial
  velocity remains almost constant over a long line of sight
  \citep[e.g.,][]{sim70,lock02}.} can also be significant. {K}inematic distance ambiguity makes the situation even worse.
{D}eviations from circular rotation will cause systematic
distortions in the kinematic distances, and hence the derived HI
distribution maps may not be reliable. On the other hand,
\citet{bur71} pointed out that the observed HI profiles can be much
better explained by velocity fields including both pure circular
rotation and streaming motions than simply assuming a pure circular
rotation, which suggested that the streaming motions predicted by
density-wave theory can mimic or mask large density differences. It
would be difficult to construct a true HI density map from the
observed HI profiles. Up to now, evidence for a spiral structure
from HI density distribution is still unclear for the inner Galaxy
\citep[][]{lock02,kk09}. For the outer Galaxy regions, however, it
is easier to map the HI distribution, because there is no kinematic
distance ambiguity. The sketch map of the main HI features obtained
by \citet{ker69} and \citet{wea70} only showed a few points of
disagreement, where the Carina Arm, Perseus Arm and
Outer Arm in the first Galactic quadrant were presented.
\citet{mdgg04} identified a new distant HI spiral arm in the fourth
Galactic quadrant, which can be traced for over 70$^\circ$ in
the $l-v$ diagram and probably is {an} extension of the
Outer Arm in the fourth Galactic quadrant. By analyzing the combined
Leiden-Argentine-Bonn all-sky HI survey data \citep[][]{kbh+05},
\citet{lbh06} constructed a perturbed surface density map of HI gas
in the outer Gala{ctic} disk. The four non-axisymmetric spiral arm
segments can be traced out to about 25{\,}kpc from the
GC.  A more recent HI map in the outer Galaxy is constructed by
identifying intensity peaks along each line of sight \citep{kpk+17}.
Beside{s} the HI emission data, evidence of spiral arms in the outer
Galaxy regions has also been seen in the HI absorption measurements
towards numerous continuum sources in the Galactic plane
\citep[e.g.,][]{sdt+07,dsg+09}.

In general, observations of{ the} HI distribution
confine the spiral morphology and kinematics of the Galaxy, yet the
spiral pattern has not been well established. Compared with young
stellar objects and molecular gas, HI gas traces a much
larger extent of the Galactic disk.

\subsection{Molecular Gas}

Molecular gas {constitutes} important components of the
interstellar medium (ISM) in the Milky Way. The highly condensed
region of molecular gas forms molecular clouds {with}
different size{s} and mass scales, and they are the
birthplace{s} of young stars. Their distributions and kinematics
represent the gas response to the Galaxy{'s} gravitational
potential, trac{ing} the gaseous spiral arms. The spiral
structure also play{s} a role in many processes involved in
molecular gas evolution, such as large-scale spiral shock,
cloud-cloud collisions, hydrodynamic instabilities, stellar feedback and magnetic fields \citep[e.g.,][]{db14}.

\begin{figure*}  %%fig2
\centering
\includegraphics[width=0.4\textwidth]{3.eps}\hspace{1cm}
\includegraphics[width=0.4\textwidth]{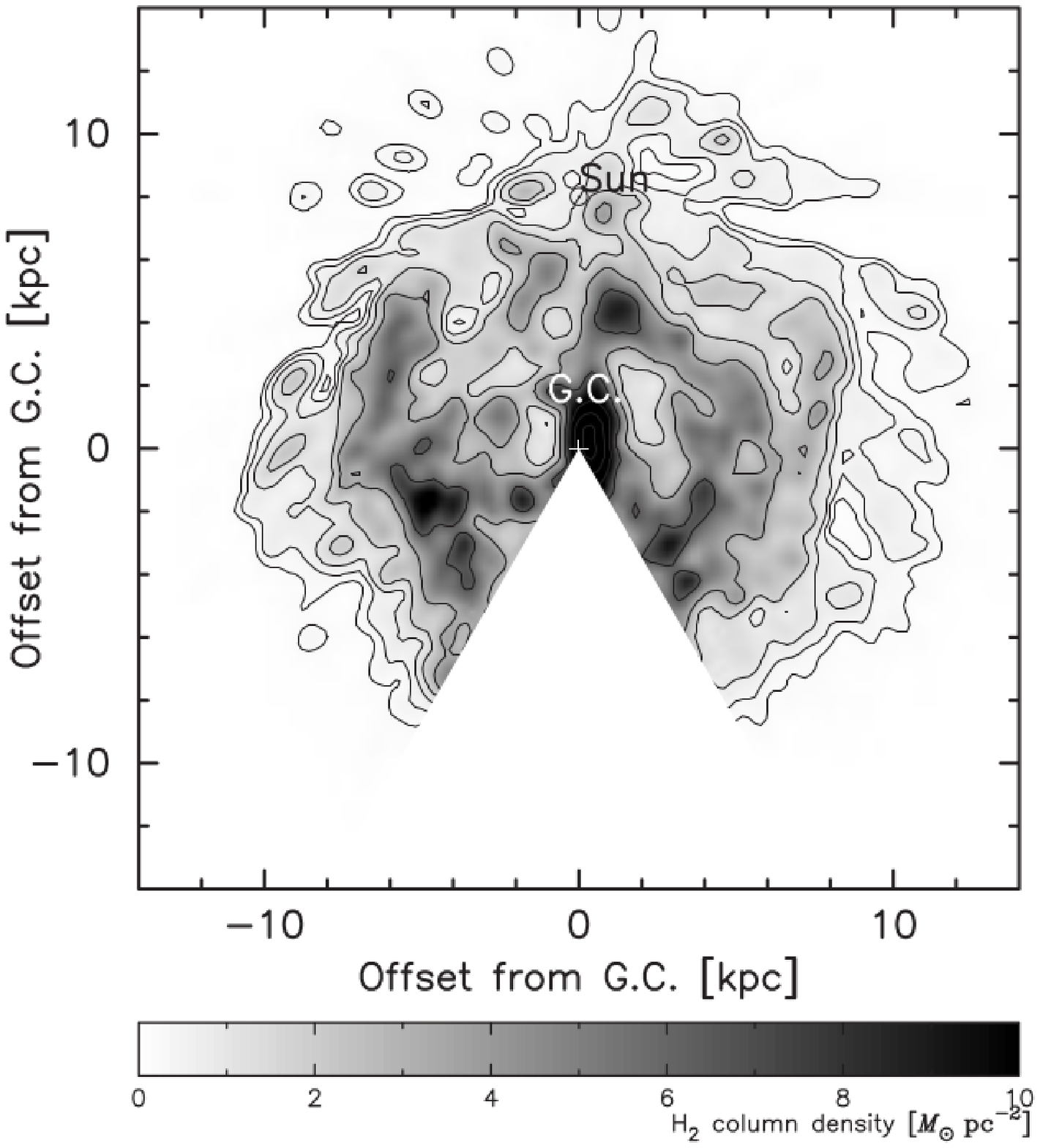}
\vspace{0.5cm}

\caption{ \baselineskip 3.8mm
 {\it Left:} Distribution of molecular
clouds in the Galactic
  disk, the data{ of which} are from \citet[][red]{mml17} and the Milky Way
    Imaging Scroll Painting project \citep[][blue]{sun15, sun17, su16,
du16, du17}. The symbol size is proportional to the mass of
  molecular clouds. The IAU standard $R_0=8.5${\,}kpc and
  $\Theta_0=220${\,}km{\,}s$^{-1}$ and the standard solar motions together
  with a flat rotation curve are adopted in deriving the kinematic
  distances. Two {\it black stars} indicate the location{s} of the Sun
  ($x=$~0.0{\,}kpc, $y=$~8.5{\,}kpc) and the GC ($x=$~0.0{\,}kpc,
  $y=$~0.0{\,}kpc). Q1 to Q4 indicate the four Galactic
  quadrants. Position uncertainties are indicated by error bars
  ({\it gray}). Galactic longitudes in degrees are marked in the plots. {\it
    Right:} Spatial distribution of molecular gas in the Galaxy
  derived by deconvolution of the CO survey data cube
  \citep[][]{dht01} by \citet{ns06}. Adapted with permission from
  Professors Nakanishi, H. \& Sofue, Y.}
\label{co}
\end{figure*}

As we reside in the Milky Way, its molecular gas content could be
surveyed in detail both with high sensitivity and resolution. Many
observational efforts have been made to construct
large-scale CO maps of the Galaxy
\citep[e.g.,][]{bg78,ccdt80,duc+87,dht01,zhang+14s, sun15, sun17,
du16, du17}. To explore the spiral structure with CO data,
three different methods are widely used:

$\bullet$ {\bf Spatial distributions of molecular clouds.} Molecular
clouds are vast assemblies of molecular gas and the
birthplace{s} of dense molecular clumps and young stars. From
the rich data set of CO surveys, a large number of
isolated molecular clouds {has been} identified, which have proven to be
good tracers of Galaxy spiral arms. Using $^{12}$CO(1$-$0) survey
data from the Columbia 1.2~m millimeter telescope in New York
\citep{ccdt80}, \citet{dect86} identified 26 GMCs. The GMC distributions in the Galactic plane
are found to resemble {those} of HII regions
\citep[][]{mdt+86}, and three segments of spiral arms, especially
the Sagittarius Arm which is clear and continuous, are delineated
in the first Galactic quadrant. Similar results were then confirmed
by \citet{sr89} with 440 molecular clouds identified from
observations of $^{12}$CO(1$-$0), and also by \citet{rjh+09} with
750 molecular clouds identified from a $^{13}$CO(1$-$0) survey using
the Five College Radio Astronomy Observatory (FCRAO) 14~m telescope
\citep[also see][]{hd15}. In the southern sky, with a {replica} of the Columbia 1.2~m millimeter telescope
in Chile, \citet{cgm+85} surveyed the southern Milky Way and
identified 37 molecular clouds in the Carina Arm. The famous
Sagittarius-Carina Arm {was} then delineated with
unprecedented clarity, which extends from the first to the fourth
Galactic quadrant{s}, more than 30\,kpc in length \citep[][]{gcbt88}.
In the fourth Galactic quadrant, some molecular clouds located in
the Crux Arm and the Norma Arm were identified by \citet{bron92}.
When the CO surveys further extended to the second and third
Galactic quadrants, more molecular clouds were identified
\citep[e.g.,][]{dig91,sod91,mab97,hcs01,gbn+14}. Some of them are
located in the far outer Galaxy
\citep[e.g.,][]{mk88,dbt90,css90,bw94,sun15,sun17}, even more than
20{\,}kpc from the Sun \citep[][]{dt11,sun17}. These
known GMCs (mass $> 10^4{\,}M_\odot$) were collected by
\citet{hh14} from literature{;} more than 1200 GMCs have
distances provided, mostly kinematic distances. More complete
catalogs of Galactic molecular clouds {were} recently
given by analyzing the classic whole-Galaxy CO maps of
\citet{dht01}. \citet{rgb+16} presented a catalog of 1064 high-mass
molecular clouds (outer Galaxy: $M_{\rm clouds} >
3\times10^3{\,}M_\odot$; inner Galaxy: $M_{\rm clouds} >
3\times10^4{\,}M_\odot$) using a dendrogram-based decomposition.
\citet{mml17} identified 8107 molecular clouds using a hierarchical
cluster identification method, shown in Figure~\ref{co}, %0000000000
including the results of the distant CO molecular clouds in the
outer and far outer Galaxy found by \citet {sun15, sun17} and \citet
{du16, du17} from the Milky Way Imaging Scroll Painting
project\footnote{\it http://www.radioast.nsdc.cn/mwisp.php}.

The major spiral arm segments
traced by GMCs can be identified, i.e., in the first Galactic quadrant:
the Scutum Arm, the Sagittarius Arm, the Perseus Arm, the Outer
Arm and even beyond the Outer Arm (Outer+1 Arm), which is
probabl{y} {an} extension of the Scutum-Crux-Centaurus Arm
\citep[][]{dt11,sun15, sun17}; in the fourth Galactic quadrant: the
Carina Arm, the Centaurus Arm and the Norma Arm. In the
{s}econd and third Galactic quadrants, consecutive
arm-like features traced by GMCs are unexplained, probably due to
the long-known velocity anomaly associated with the Perseus Arm
\citep[e.g.,][]{fc10}. Although molecular clouds are good tracers of
the global picture of spiral structure, only kinematic distances are
available for the majority of known molecular clouds,
which have large errors \citep[e.g., see][]{rb18}. Large distance
uncertainties of molecular clouds precluded making a true
three-dimensional (3D) map of the Milky Way with sufficient accuracy
to trace its spiral structure.

$\bullet$ {\bf Deconvolution of the CO survey data cube.} Rather
than through the identifications of molecular clouds, the spatial
distributions of molecular gas in the Milky Way could be constructed
directly from the CO survey data cube, which are instructive to
understand the global spiral structure, and also highly useful for
understanding diffuse Galactic gamma-ray emission
\citep[e.g.,][]{hbc+97} and the propagation and properties of cosmic
rays \citep[e.g.,][]{jpm18}. To interpret the observed properties of
diffuse gamma-ray emissions, \citet{hbc+97} constructed a surface
density map of molecular gas from CO surveys \citep[][]{duc+87}.
Based on the whole-Galaxy CO maps of \citet{dht01}, \citet{ns06}
created a 3D distribution map of molecular gas throughout the
Galactic plane, and some concentrated area{s} of molecular gas are
visible and consistent with the features shown in the molecular
cloud map (see Fig.~\ref{co}), but not as clear as those traced by
HII regions. The gas concentrations probably are related to the
major spiral arm segments as shown in figure~14 of \citet{ns06}.
After that, \citet{ns16} presented a new combination of HI
and H$_2$ surface density maps with similar methods. \citet{peb08}
used a gas-flow model to derive a model of the spatial
distribution of molecular gas in the Milky Way from the CO survey
maps of \citet{dht01}, rather than simply assuming a pure circular
rotation picture. They found a concentration of mass along the
Galactic bar, and at the ends of the bar, two spiral arms emerge.
{However,} the evidences for other spiral arms are not
strong as shown from the deconvolution map of surface density
\citep[fig.~6
  of][]{peb08}.

The global 3D distributions of molecular gas could be re-constructed
by deconvolution of the survey data cube $T_b(l,b,v)$ of CO. This
method depends on the adopted model of the Galaxy velocity field
(e.g., circular or non-circular), which is basic to derive
the kinematic distances of molecular gas. It also depends on the
near-and-far or kinematic distance ambiguity. From the
current results, spiral arm features shown by this method are
not clear or continuous. It seems impossible to determine the global
spiral structure from this method alone.

$\bullet$ {\bf Modeling the observed longitude-velocity maps of CO.}
From the CO survey ($l$-$b$-$v$) data toward the Galaxy, the
longitude-velocity ($l-v$) diagram can be created by integrating
emission over latitude \citep[e.g.,][]{duc+87,dht01}. The
large-scale distributions and kinematics of molecular
clouds in the Galaxy are embedded in the emission features shown in
the $l-v$ map \citep[fig.~3
  of ][]{dht01}, which indicate the concentrations of stars and
interstellar gas, tracing some remarkable structures such as the
spiral arms, the 3-kpc arms, the Galactic Molecular Ring and arm
tangencies. To interpret and transform the observed $l-v$ map to a 3D
distribution of gas, a model of gas flow is required. Many
numerical efforts have been made \citep[e.g., see][]{db14,pdap14}.
\citet{eg99} presented a model for gas dynamics in the
inner Galactic plane{.} {T}heir best model
leads to a four-armed spiral structure and reproduces the observed
directions towards five arm tangencies. \citet{fux99} modeled gas
dynamics in the Galactic disc with a 3D $N$-body
simulation{.} {T}he gas flow in the model can
reproduce some major features in the CO $l-v$ map, but only at
specific times, suggesting a transient nature of the
Galaxy{'s} spiral arms. \citet{rc08} modeled the gas flow by including
the nuclear bar constrained by 2MASS data{.}
{T}heir simulations reproduced the major spiral arms, the
near and far 3-kpc arms, and interpreted the Galactic Molecular Ring
as the inner parts of spiral arms rather than an actual ring.
\citet{bsw10} used a{n} $N$-body and hydrodynamical simulation
to model the CO $l-v$ map, in which multi-phase ISM, star
formation and supernova feedback were considered. They qualitatively
reproduced the large-scale emission features of the CO $l-v$ diagram
and also clumpy structures. In a face-on view of their best gas flow
model, the Milky Way looks more like a multiple-arm or
flocculent{ galaxy} rather than a grand design
spiral galaxy. \citet{pdap14} used smoothed particle hydrodynamics
to simulate gas flow in the Milky Way by assuming a grand design
spiral (four-armed or two-armed){.} {T}hey
found that it is possible to reproduce the major features shown in
the $l-v$ map of CO, but neither four-armed nor two-armed models can
reproduce all of the observed features simultaneously. Then,
\citet{pdab15} took a different approach by modeling the stellar
distribution with many discrete $N$-body particles rather than a
continuous gravitational potential. Their best fitted models can
match the observed CO $l-v$ map much better than previous work and
favor a four-armed structure, but the spiral arms are dynamic and
transient. Generally, the precise global spiral pattern and the
formation mechanism cannot be uniquely determined from current
gas flow models of the Milky Way.

\subsection{Open Clusters}
\label{sect:Open}

\begin{figure*}  %%fig3
\centering\includegraphics[width=0.4\textwidth]{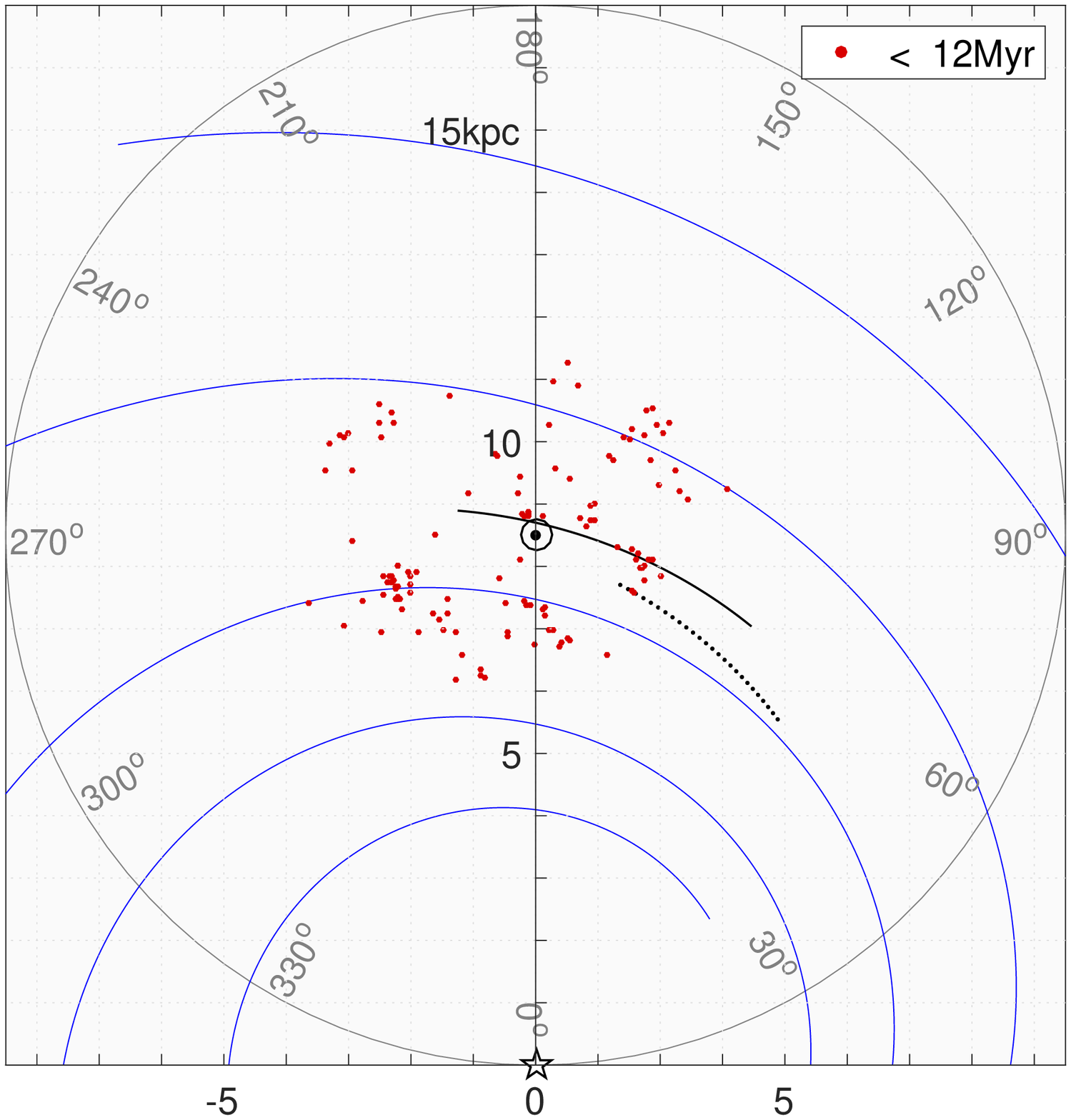}\hspace{1cm}
\includegraphics[width=0.4\textwidth]{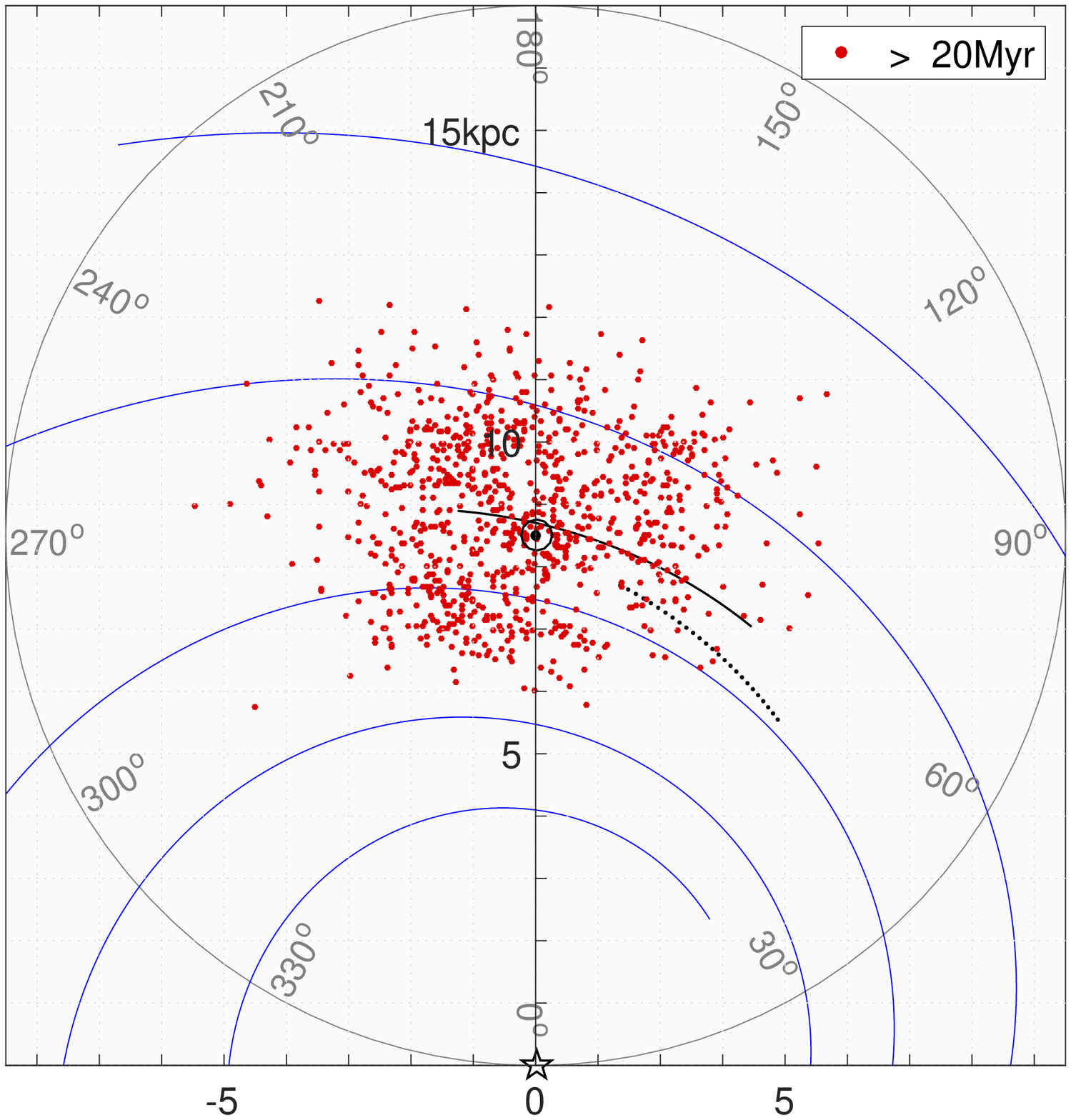}
\vspace{0.5cm}

\caption{\baselineskip 3.8mm {\it Left:} Distribution of young open
clusters with age $<$
  12{\,}Myr. {\it Right:} Distribution of older open clusters with age
  $>$ 20{\,}Myr. Two {\it black stars} indicate the location{s} of the Sun
  ($x=$~0.0{\,}kpc, $y=$~8.5{\,}kpc) and the GC ($x=$~0.0{\,}kpc,
  $y=$~0.0{\,}kpc). Galactic longitudes in degrees are also marked in the
  plots. The cluster data are taken from \citet[]{he18}. The
  background is the spiral arm model obtained by \citet{hh14} through
  fitting the Galactic distributions of known HII regions with a
  polynomial-logarithmic spiral arm model, except the Local Arm, whose
  parameters are adopted from the recent work of \citet{xrd+16}. From
  top to bottom, they are the Outer Arm, the Perseus Arm, the Local
  Arm, the Sagittarius-Carina Arm, the Scutum-Crux-Centaurus Arm and the
  Norma Arm.}
\label{oc}
\end{figure*}

The open clusters (OCs) have a wide range of ages, from a few
million years to more than ten billion years. The young OCs are too
young to migrate far from their birth locations. For old OCs, they
gradually drift away from their birthplace and move to inter-arm
regions. \citet{bec63,bec64} first studied the relation between OCs and spiral arms using a sample of 156 OCs with photometric
distances. He pointed out that the distribution of OCs with the
earliest spectral type between O and B2 probably followed three
spiral arm segments in the vicinity of the Sun, and resembled the
distribution of nearby HII regions, while the distribution of older
OCs with spectral type between B3 and F did not indicate spiral arm
segments and seemed to be random. These conclusions were confirmed by
\citet{bf70} and \citet{fb79} with larger samples of OCs. With a
sample of 212 OCs, \citet{dl05} showed that the OCs with ages up to
about 1.2$\times 10^{7}${\,}yr remain in parts of the
Perseus Arm, the Local Arm and the Sagittarius-Carina Arm; those
with ages $\sim$20{\,}Myr are leaving the spiral arms and
filling the interarm regions; for clusters older than
30~Myr,{ the} spiral or clumpy-like structure has disappeared in their
distribution. Therefore, it is generally believed that young OCs are
tracers of Gala{ctic} spiral arms. However, from the distributions of
about 120 open clusters with age $<$ 10$^{7.5}${\,}yr,
\citet{lyng82} suggested these features seem like three clumpy-like
concentrations or complexes, rather than associations to extended
spiral arms. \citet{ja82} and \citet{jtl88} independently obtained a
similar conclusion from a larger sample ($>$ 400 young OCs). They
suggested that the distributions of OCs look like some clumps or
short arm segments, {with }no spiral pattern at all. At present,
the number of known open clusters has significantly
increased to more than 3000 \citep[see
e.g.,][]{daml02,kps+13,skp+14,lp17,
  he18}. As shown in Figure~\ref{oc}, a majority of them {is} located
within about 3{\,}kpc from th{e} Sun. The 126 OCs
with age less than 12{\,}Myr are concentrated in parts of
the Perseus Arm, the Local Arm and the Sagittarius-Carina
Arm. On the contrary, the old OCs seem to {be }distribute{d}
randomly. Due to the extinction of interstellar dust, it would be
difficult to identify distant young OCs. At present, it seems not
possible to infer the global spiral structure of the Milky Way from
OCs alone.

\subsection{Arm Tangencies in the Inner Galaxy}

\begin{figure*}  %%fig4
\centering \includegraphics[width=6cm]{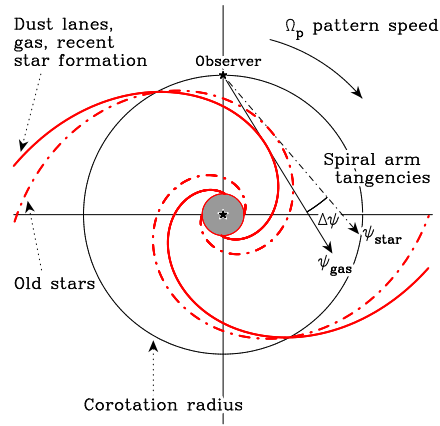}

\caption{\baselineskip 3.8mm  Schematic of spiral arm tangencies for
gas ($\psi_{\rm gas}$) and
  old stars ($\psi_{\rm star}$), which are indicated by the density peaks
  and/or irregular terminal velocities near the tangency points of
  ideal spiral arms{;} also shown are the relative position{s} between the
  spiral arms traced by gas and old stellar components according to
  the predictions of quasi-stationary density wave theory \citep[e.g.,][]{rob69}. Adapted with permission from
  \citet{hh15}.}
\label{show}
\end{figure*}

As shown by face-on spiral galaxies (e.g., M101, Fig.~\ref{M101}),
%000000000000
 spiral arms appear to be long and thin features in the
distributions of interstellar gas and stars. These large-scale
structures originate near the galaxy center and extend to the far outer edge,
normally approximated by logarithmic form
\citep[e.g.,][]{hr15}. As shown in Figure~\ref{show}, %000000000000
tangencies of spiral arms are expected for an observer inside the
host galaxy. As to our Milky Way, the arm tangencies have long been
known as one of the best{ pieces of} evidence {for}
spiral arms in the inner Galaxy, and provide useful observational
constraints to spiral structure models
\citep[e.g.,][]{bs70,gg76,ol93,eg99,ne2001,rus03,ben08,hh14}. In
addition, the possible displacements of arm tangencies for different
components (e.g., gas and stars) can be used to constrain the
formation mechanisms of spiral arms in the Milky Way
\citep[e.g.,][]{val14,hh15}.

\begin{table*}  %%table1
  \caption{\baselineskip 3.8mm Galactic longitudes of spiral arm tangencies for old stars
    and interstellar gas, identified from multiwavelength
    observational data \citep[see][]{hh15}.}
\begin{center}
%\begin{threeparttable}

\renewcommand\baselinestretch{1.4}
\fns\tabcolsep 3mm
\begin{tabular}{lccccccc}
  \hline\noalign{\smallskip}
Component       & Near 3{\,}kpc North & Scutum & Sagittarius & Carina & Centaurus & Norma & Near 3{\,}kpc South  \\
  & $(^{\circ})$  & $(^{\circ})$ & $(^{\circ})$ & $(^{\circ})$ & $(^{\circ})$& $(^{\circ})$ & $(^{\circ})$   \\
  \hline\noalign{\smallskip}
Old stars            & 27.0        &  32.6 & 55.0$^*$  &  --     & 307.5         & --     & 338.3       \\
Interstellar gas     & 24.4$^*$    &  30.7 & 49.4      &283.8    & 305.5, 311.2  & 328.1  & 337.0$^*$   \\
  \hline\noalign{\smallskip}
\end{tabular}
%\begin{tablenotes}
\parbox{155mm}
{ \footnotesize
Notes: $^*$, the measured arm tangencies have
  lower confidence, as the corresponding local maxima features {are} not
  present in all of the investigated dataset{s}.}
%\end{tablenotes}
%\end{threeparttable}
\end{center}
\label{tan}
\end{table*}

Arm tangencies have been identified by a number of research works
from the survey of molecular gas, atomic gas, ionized gas,{ and}
young and old stellar components in the Galactic disk. A recent
compilation was given by \citet{val14,val16}, which ignores the different
definitions of arm tangencies in references. The derived arm
tangencies by different definitions may be different by about half
of the arm width even from the same observation data. To properly
measure the arm tangencies, a consistent definition and
multi-wavelength survey data should be considered together.

$\bullet$ For baryons in the {ISM} (e.g., ionized gas, atomic gas{ and}
molecular gas), the tangencies of spiral arms have been measured:

(1) by the bump features or condensed emissions appearing in
the ($l,v$) diagrams of HI \citep[e.g.,][]{bs70,rmw+84}, CO \citep[e.g.,][]{gcb+87,eg99},
HII regions \citep[e.g.,][]{lock89,eg99} and methanol masers
\citep[e.g.,][]{cfg+11};

(2) by the excess velocity features shown in the terminal velocity
curve of HI \citep[e.g.,][]{bur71,sha72,md07}, CO
\citep[e.g.,][]{cle85,amb90} and HII regions
\citep[e.g.,][]{rbcw86}, which were interpreted as the result of
streaming motions along the spiral arms, and/or large internal
motions of the cloud complexes;

(3) by the solid-body like kinematic features in the smoothed rotation
curve of CO \citep[][]{lbcm06}, which meet the peaks in the gas
surface density curve and the valleys in the shear and vorticity curve;

(4) by the local peaks or enhancement in the integrated emissions of
CO, HI, {radio recombination lines (}RRLs{)}
or thermal radio continuum over latitude/velocity with Galactic
longitude \citep[e.g.,][]{bkb85,bact89,swh10}; and by
local maxima in the integrated number counts of HII regions,
6.7\,GHz methanol masers, dense molecular clumps or pulsars,
over Galactic latitude and plotted against longitude
\citep[e.g.,][]{ne2001,hh15};

(5) from the best spiral arm models fitted to the projected
distributions of spiral tracers (such as HI, molecular clouds, HII
regions{ and} HMSFR masers) in the Galactic plane
\citep[e.g.,][]{rus03,hhs09}, or from the models fitted to the
($l,v$) diagram \citep[e.g.,][]{eg99}. The derived arm tangencies
are suggested to be less confident than the above four methods.

$\bullet$ As to the stellar components in the inner Galaxy, it is
still difficult to measure radial velocities for a large number of
distant stars due to interstellar dust extinction.  The arm
tangencies were commonly identified by local maxima shown in the
integrated number counts of near-infrared (NIR) or far-infrared
(FIR) point sources{,} or the integrated NIR or FIR emissions
against Galactic longitudes \citep[e.g.,][]{dri00,ds01,ben08}. The
interstellar extinction seems {t}o not significantly
influence the measured longitudes of arm tangencies from the survey
data of old stars \citep[][]{ben08,hh15}. It should be mentioned
that the identified arm tangencies by measuring the local maxima
deviate from the true density maxima of matter (gas or stars) near
the arm tangencies by shifting to the inner side due to{ the effect of} integration along the line of sight. The discrepancy is
probably small, i.e., less than about 1$^\circ\sim2^\circ$ in
longitude \citep[][]{dri00,hh15}.

We emphasize that there are pitfalls in the above methods used to
identify arm tangencies.  {V}elocity crowding,
including streaming motions \citep[e.g.,][]{bur73}, can
result in ``bump'' features in the ($l,v$) diagram. The
concentrations of individual clouds, star-forming regions or old
stars can produce local maxima in the integrated number
count plots and the integrated emission plots against Galactic
longitudes. The nearby clouds, star-forming regions
or old stars could be wrongly recognized as
longitud{inal} concentrations of farther objects and then
be misinterpreted as arm tangencies. In addition, observations of interstellar
gas or old stars towards arm tangencies could be complicated by dust
extinction effects. Observations of small distant objects suffer
from {the }beam dilution effect. In order to properly identify
arm tangencies for a better understanding of the Milky Way{'s}
spiral structure, a consistent definition and
multiwavelength surveys {of} different Galactic
components should be considered together, because the problems
discussed above may {be }present in one or two data sets,{
but} not in all datasets. A careful re-evaluation of arm tangencies
with more survey data of interstellar gas and stars would be useful.

Such kind of work has recently been done with the method of
identifying local maxima in the longitude plots of source number
counts for GLIMPSE/2MASS sources, HII regions, 6.7\,GHz methanol
masers, dense clumps and in the plots of integrated emissions
for RRLs, HI, $^{12}$CO and $^{13}$CO \citep[][]{hh15}. The arm
tangencies identified for different gas components in the ISM, i.e.,
HII regions, methanol masers, CO gas, dense molecular clumps
and HI gas, appear{ at} nearly the same Galactic longitudes. The
arm tangencies for GLIMPSE and 2MASS old stars also appear
at nearly the same longitudes. The results are summarized in
Table~\ref{tan} for comparisons with spiral structure
models of the Milky Way. By using other definitions, e.g., excess
velocity features in the terminal velocity curve, a systematic
re-evaluation of arm tangencies with modern survey data has not yet
been done. Such investigations may provide some insight into
uncovering the possible displacements of arm tangencies between
different gas components, which was found by \citet{val14}, but not
confirmed by \citet{hh15}. In addition, the tangencies for some arm
segments are still uncertain, e.g., the Sagittarius Arm tangency for
old stars, the Near and Far 3-kpc Arm tangencies for
gas (see Table~\ref{tan}), as the corresponding
local maxima features{,} were not found in all of the studied data
sets, and deserve more attention with survey data of stars and gas
in the near future.

\section{VLBI {and} {{\it Gaia}} astrometry}
Over the past decade, the astrometric accuracy of {VLBI} has improved dramatically.
{P}ioneer{ing} work measured the trigonometric
parallaxes and proper motions to masers associated with
{HMSFR} W3OH and obtained an
accuracy of 10{\,}$\upmu$as \citep{xrzm06}, allowing us
to perform precise distance measurements to objects at the GC and
beyond, extending to the outer edge of the Galaxy. This is a
landmark in this field \citep{binney06,
  cas+12}. Currently{,} relative positions between sources separated by
about one degree are being measured with accuracies of a few
$\upmu$as \citep{Honma:07, Reid:09a, Zhang+13, srd+17}. With this
accuracy, trigonometric parallaxes can be obtained accurately
throughout the Milky Way.  These techniques have been applied to
VLBI networks like the NRAO Very Long Baseline Array (VLBA) in the
US{A}, the European VLBI Network (EVN)\ in Europe and China, or
the VLBI Exploration of Radio Astrometry (VERA) array in Japan
\citep{reidhonma2014}.

The astrometric satellite {{\it Gaia}} is expected to
significantly augment our knowledge about Galactic
structure and space motions. With the most accurate astrometric
parameters of the youngest O stars and masers, for the first time,
the spiral structure in all four quadrants{ has been}
delineated clearly in unprecedented detail \citep{Xu+18}. The
revealed Galactic spiral patterns make a clear sketch of nearby
spiral arms, especially in the fourth quadrant where maser parallax
measurements are absent.

In addition to distances, {{\it Gaia}} also
yield{s} excellent measurements of secular proper motions, with
accuracies of $\approx$1{\,}\kms. Combining radial
velocities with proper motions (and distances) yields full 3D
velocities, relative to the motion of the Sun. Thus, through this
measurement, one may also be able to determine the full kinematics
in the Milky Way, which can accurately define {its
associated} rotation curve
\citep{Brunthale11, Honma:12, Reid:14}. Therefore, the current
astrometry can provide an excellent opportunity to map our Galaxy in
great detail, {yielding} the precise geometry, Galactic fundamental
parameters and 3D velocity field.

In this section{,} we review progress on
spiral structures of the Milky Way made during the past decade
relying on VLBI {and} {{\it
Gaia}} astrometry. Meanwhile, we present our own latest research
results from astrometric measurements.

\subsection{New Galactic Spiral Arms}

Here we display the spiral structure revealed by O stars from
{{\it Gaia}} DR2 and VLBI maser parallax measurements.
Because some sources have considerable uncertainties (typically more
than 20\% and a few even more than 30\%) in the{ir} parallaxes,
which are comparable to the size of spacing between arms, only those with distance accuracies better than 15\% are
adopted{.} {C}ollectively{,} 102 masers
\citep{Ando:11,Asaki:10,Bartkiewicz:08,Brunthaler:09,Choi:08,Choi:14,hbm+06,hbm+09,hcr+15,Hirta:08,Honma:07,Honma:11,Immer:13,Kim:08,Kura:11,
  Menten:07,Moell:09,Mos:09,Mos:11,Nagayama:11,Nii:11,Oh:10,Reid:09a,Reid:09c,Rygl:10,Rygl:12,Sandstrom:07,sanna09,Sanna:12,Sanna:14,Sato:08,
  Sato:10a,Sato:10b,Sato:14,Shio:11,Wu:14,xrzm06,Xu+09,Xu+11,Xu+13,Zhang+09,Zhang+12a,Zhang+12b,Zhang+13,Zhang+14}
and 635 O stars \citep{Xu+18}{ are} listed in
Tables~\ref{tab:paraM} {and} ~\ref{tab:paraO}{ respectively}.

\begin{table*}  %%table2
\centering
\begin{minipage}{8cm}
    \caption{Parallaxes and Proper Motions of Masers\label{tab:paraM}}
\end{minipage}
%\begin{threeparttable}

\renewcommand\baselinestretch{1.4}
\fns\tabcolsep 3mm
    \begin{tabular}{lcccccccc}
  \hline\noalign{\smallskip}
        Name &  R.A.&Dec. & $\pi$ &$\mu_{x}$ & $\mu_{y}$
        &$\upsilon_{\rm LSR}$ &Spiral \\
          & $(\deg)$& $(\deg )$& (mas) & (mas{\,}yr$^{-1}$)&(mas{\,}yr$^{-1}$)
        & (km{\,}s$^{-1}$)& Arm\\
(1)&(2)&(3)&(4)&(5)&(6)&(7)&(8)\\
  \hline\noalign{\smallskip}
L 1287&09.1975&63.4841&1.077 $\pm$ 0.039&$-$0.86 $\pm$ 0.11&$-$2.29$\pm$0.56&$-$23 $\pm$ 5&Loc\\
G122.01-07.08&11.2433&55.7799&0.460 $\pm$ 0.020&$-$3.70 $\pm$ 0.50&$-$1.25$\pm$0.50&$-$50 $\pm$ 5&Per\\
G123.06-06.30&13.1008&56.5620&0.421 $\pm$ 0.022&$-$2.69 $\pm$ 0.31&$-$1.77$\pm$0.29&$-$29 $\pm$ 3&Per\\
G123.06-06.30&13.1030&56.5640&0.355 $\pm$ 0.030&$-$2.79 $\pm$ 0.62&$-$2.14$\pm$0.70&$-$30 $\pm$ 5&Per\\
G134.62-02.19&35.7155&58.5865&0.413 $\pm$ 0.017&$-$0.49 $\pm$ 0.35&$-$1.19$\pm$0.33&$-$39 $\pm$ 5&Per\\
        ...\\
  \hline\noalign{\smallskip}
    \end{tabular}
%\begin{tablenotes}
\parbox{140mm}
{\baselineskip 3.8mm Notes: Column~(1) gives the Galactic
          source name. Columns~(2) and (3) are Right Ascension and
          Declination (J2000), respectively. Columns~(4) to (6) give the
          parallax and proper motion in the eastward ($\mu_x=\ura
          \cos{\delta}$) and northward directions ($\mu_y=\udec$),
          respectively. Column~(7) lists local standard of rest
          (LSR)
          velocity. The same as \citet{Reid:14}, Column~(8) indicates
          the spiral arm in which it resides. The full table is
          available in the electronic attachment ({\it http://www.raa-journal.org/docs/Supp/ms4260table2.dat}).}
%\end{tablenotes}
%\end{threeparttable}
\end{table*}

\begin{figure*}  %%fig5
\centering
\includegraphics[width=10cm]{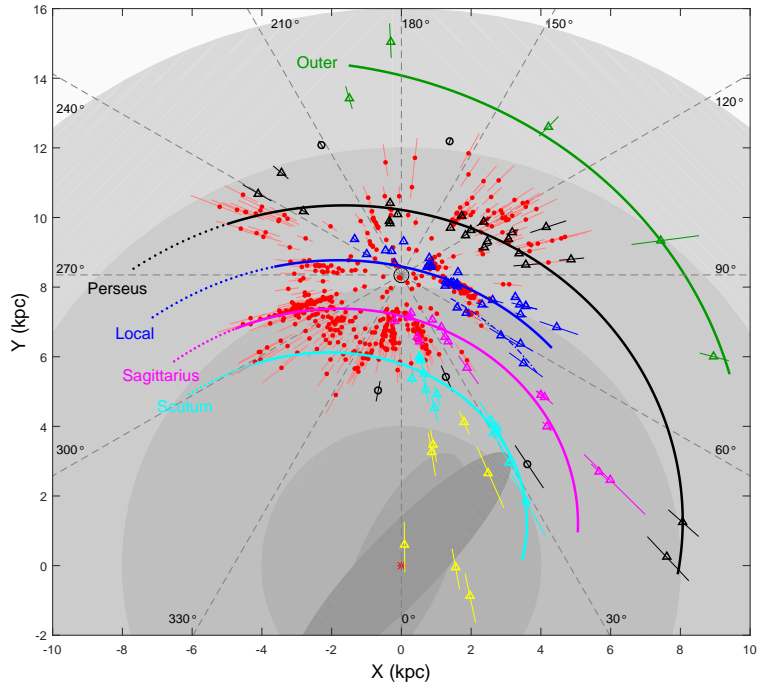}

\caption{ \baselineskip 3.8mm Up to date face-on view of the spiral
arms determined from
  parallaxes of masers ({\it triangles}) and O stars ({\it red circles}).
  The formal parallax uncertainties of the sources shown here are better
  than 15\%. {\it {S}olid curved lines} depict the log-periodic spiral
  fitting, while {\it dotted lines} are generated by extrapolating the
  log-periodic spirals. Here \Ro = 8.35\,kpc (see {Sect.~3.2}).
\label{xy-plot}
}
\end{figure*}

\begin{table*}  %%table3
\centering
\begin{minipage}{8cm}
\caption{Parallaxes and Proper Motions of O Stars\label{tab:paraO}}
\end{minipage}
%\begin{threeparttable}
\fns\tabcolsep 1.5mm
\begin{tabular}{lcccccccccc}
\hline\noalign{\smallskip} Name & {{\it Gaia}} DR2 ID
& R.A.&Dec. & $\pi$ &$\mu_{x}$ & $\mu_{y}$
&$\upsilon_{\rm LSR}$ &Spectral&Spiral \\
&  & $(\deg)$& $(\deg)$& (mas) & (mas{\,}yr$^{-1}$)&(mas{\,}yr$^{-1}$)
& (km{\,}s$^{-1}$)& Type&Arm\\
(1)&(2)&(3)&(4)&(5)&(6)&(7)&(8)&(9)&(10)\\
\hline\noalign{\smallskip}
ALS 13375&528594342521399168&0.4453 &67.5070 &1.016 $\pm$0.031 &$-$1.57 $\pm$0.04 &$-$1.77 $\pm$0.04 &&O9.5V&Loc\\
ALS 13379&528570015826682496&0.5429 &67.4089 &0.928 $\pm$0.035 &$-$1.61 $\pm$0.05 &$-$2.04 $\pm$0.05 &$-$0.6&O7V&Loc\\
ALS 6006&429470895385555456&0.9896 &61.1036 &0.289 $\pm$0.032 &$-$0.85 $\pm$0.05 &$-$1.74 $\pm$0.05 &$-$20.2 $\pm$2&O9.7Iab&Per\\
ALS 6009&429927879906030336&1.0158 &62.2219 &0.295 $\pm$0.037 &$-$1.40 $\pm$0.06 &$-$1.72 $\pm$0.05 &$-$36.6 $\pm$0.3&O8Iabf&Per\\
ALS 6014&528409143531333376&1.0673 &66.3491 &0.917 $\pm$0.038 &$-$1.15 $\pm$0.05 &$-$4.19 $\pm$0.05 &&O9&Loc\\
...\\
\hline\noalign{\smallskip}
\end{tabular}
\parbox{155mm}
{ \baselineskip 3.8mm
%\begin{tablenotes}
\footnotesize
%\item[]
Notes: Column (1) is the name of Alma luminous
  star (ALS); Column (2) is the unique source identifier in {{\it Gaia}}
  DR2. Columns (3) and (4) are Barycentric Right Ascension (R.A.) and
  Declination (Dec.), respectively. Columns 5 to 7 give the parallax
  and proper motion in the eastward ($\mu_x=\ura \cos{\delta}$) and
  northward directions ($\mu_y=\udec$), respectively. Column 8 lists
  local standard of rest (LSR) velocity. Column 9 is the{ specific} spectral
  {sub}type
  of the O stars. Column 10 indicates the spiral arm in which it
  resides. In {{\it Gaia}} DR2, {t}he reference epoch is J2015.5. The full table
  is available in the electronic attachment ({\it (http://www.raa-journal.org/docs/Supp/ms4260table3.dat}).}
%\end{tablenotes}
%\end{threeparttable}
\end{table*}

In order to purify the O star sample that {is} truly
capable of tracing spiral arms, we further eliminated those with
peculiar motions in the direction perpendicular to{
the} Galactic plane larger than 20{\,}\kms\ and {with
}a 3D velocity of more than 60{\,}\kms, resulting{
in} a total of 583 O stars. They are the youngest stars and their
peculiar motions are relatively small, which means they are supposed
to be located near their birthplaces. With a typical lifetime of 3
Myr~\citep{Weidner10}, O5.0
\uppercase\expandafter{\romannumeral3} stars move $\sim$0.2\,kpc
from their birth places at a speed of 60{\,}\kms. Because
the width of spiral arms neighboring the Sun ranges from 0.2 to
0.4\,kpc \citep{Reid:14} and the lines of sight usually are not
perpendicular to spiral arms, the remaining O stars are guaranteed
to be located in their spiral arms.

As shown in Figure~\ref{xy-plot}, %00000000000000
 the conjunctions of
VLBI and {{\it Gaia}} parallax results are distributed
in strip{s} and clump{s}, meaning that they trace spiral arms, while the
sources are relatively sparse indicating the gap between spiral
arms. For the first time, these data fill in the whole sky,
especially the previous gap from 240$^{\circ}$ to 360$^{\circ}$
along Galactic longitude. As expected, most O stars {are }gather{ed}
around the Sun within a radius of $\sim$3\,kpc, while the masers,
despite a relatively small number, {are
}distribute{d} much{ more} wide{ly} than the O stars, even over 10\,kpc.
In general, they reveal a clear spiral pattern, consisting of five
obvious spiral arm segments. From top to bottom, they are part of
the Outer Arm, the Peruses Arm, the Local Arm, the Sagittarius Arm
and the Scutum Arm. These measurements strongly support the
existences of spiral arms in the Milky Way {G}alaxy.

Masers are assigned to arms based on their coincidence in Galactic
longitude and velocities in the Local Standard of Rest frame ($V_{\rm LSR}$)
with CO and HI $l-v$ emission features \citep{Reid:14,xrd+16},
while for O stars, because of their large peculiar motions, the
method based on $\log(r)-\theta$ information is the same as \citet{hh14} for HII regions and GMCs. Here, $r$ represents the
distance to the GC, and $\theta$ starts from the positive $x$-axis
and increases counterclockwise.

Although there are only four maser sources in a long longitude
distribution, from $l$ $\approx$ 74\deg\ to 190\deg, they outline an
arc shape, i.e., part of{ an} arm segment in the Outer
Arm{.} The Peruses Arm has a large number of
HMSFRs. Between $l$
$\approx$ 90\deg\ and 210\deg, masers and O stars mix well, firmly
tracing the arm segment. On one end{,} the maser data
extend to $l$ $\approx$ 45\deg, but{ there is} a lack of masers
and O stars between $l\approx45^\circ$ and $90^\circ$ in this arm.
On the other end, the O stars extend the arm segment to
$l\approx255^\circ${.} The Local Arm is the nearest
spiral arm to the Sun. Considering all relevant optical and radio
data available at the time, \citet{gg76} concluded that the Local
Arm was a ``spur'' or a secondary spiral feature, because the
density of star forming regions appeared to be significantly less
than that of other major arms in the Milky Way. However,
\citet{Xu+13} found a larger number of star forming regions in this
arm, some of which{ were} thought to be in the Perseus Arm,
suggesting that the Local Arm is a major spiral structure.  The
Local Arm stretches approximately from $l$ $\approx$ 70\deg, past
the Sun, upward slightly and at $l$ $\approx$ 240\deg\ bend{s}
down to $l$ over 270\deg\ with one branch (see shadow area) touching
the Perseus Arm at around $l$ $\approx$ 200\deg.\ In spite of a higher abundance of high-mass stars, this branch
resembles the spur that links the Local {A}rm and the
Sagittarius Arm discovered by \citet{xrd+16}{.} Unlike
{the }vast sources in the {Perseus} and Local
{A}rms, there are not too many masers located in the
Sagittarius Arm. Most of the masers {are }concentrate{d}
between $l\approx0\deg$ and 30\deg. Only a few masers extend to $l\approx 45$\deg.\
On the other end, O stars stretch into the fourth quadrant
at about $l\approx 285\deg${.} {M}asers
located in the Scutum Arm are confined between $l\approx 0$\deg\ and
30\deg. On the other end, the O stars may extend the Scutum Arm from
$l\approx 0$\deg\ to $l\approx 300$\deg\ into the fourth quadrant.

\begin{figure*}  %%fig6
\centering
\includegraphics[width=10cm]{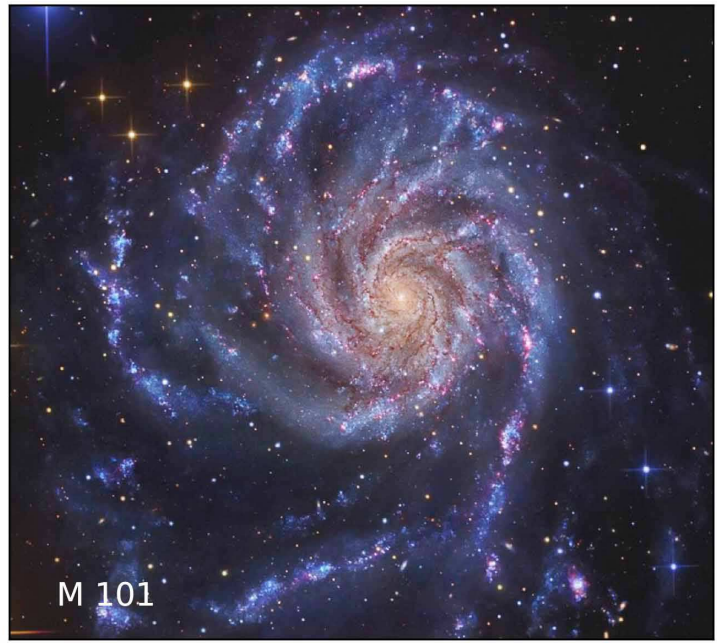}

\caption{\baselineskip 3.8mm The galaxy M\,{\,}101. The Milky
Way is likely to be this{ type of} galaxy
  with abundant branches/spurs. Adapted with permission from
  Dr. R. Jay GaBany.
\label{M101}
}
\end{figure*}

Figure~\ref{xy-plot} shows many more possible branches/spurs.
Besides the Local spur that links the Local Arm and the Sagittarius
Arm \citep{xrd+16} and links the Local Arm and the Perseus Arm at
around $l\approx 200\deg$, other spurs are identifiable. Between the
Sagittarius Arm and Scutum Arm there are a few possible
spurs that cannot be conclusively confirmed due to large distance
uncertainties. Near $l\approx 280\deg$, it looks
like one branch of the Sagittarius Arm going upward and
connecting{ to} the Local Arm. Although current parallax data are inadequate to
clearly describe the entire Galactic structure, based on present
results, our Galaxy may have many sub-structures in addition to
its major arms. This suggests that our Galaxy is
quite different from a pure grand design spiral galaxy with
well-defined, two- or four-major arms{ being the}
domina{n}t, such as M51, although a pure
grand design morphology is more popular. Our Milky Way largely resembles an
external galaxy,
the Pinwheel Galaxy (M 101) (Fig.~\ref{M101}). %0000000000000

The pitch angle is an indicator of the tightness of spiral arms.
Usually, a logarithmic model is used to fit pitch angles because
spiral arms of galaxies crudely agree with a logarithmic form
\citep{Kennicutt:81,hr15}. Following the method of
{\cite{Xu+13}}, we fitted masers and O stars
together {with} arm segments, adopting a log-periodic
spiral defined by
\begin{equation}\label{logic}  %%eq1
\ln(R/R{\rm_{ref}})=-(\beta-\beta{\rm_{ref}})\tan\psi\;,
\end{equation}
where $R$ is the Galactocentric radius at a Galactocentric azimuth
$\beta$ (defined as 0 toward the Sun and increasing clockwise) for
an arm with radius $R{\rm_{ref}}$ at reference azimuth
$\beta{\rm_{ref}}$ and pitch angle $\psi$. To search for the
optimized values of each parameter, we minimized the factor
\begin{equation}\label{fitz}  %%eq2
Z=\dfrac{1}{\sum W_{i}}\sum W_{i}\sqrt{(x_{i}-x_{t})^{2}+(y_{i}-y_{t})^{2}}\;,
\end{equation}
where $W_{i}$ is the weight. We simply assigned the weight factor
$W_{o}$ = 1 for the O stars{ and} the weight factor
$W_{m} $=10 for masers, because the amount of masers is much smaller
but they {are }distribute{d} much{ more}
wide{ly}; $x_{i}$ and $y_{i}$ are the Cartesian
coordinates of a spiral tracer; $x_{t}$ and $y_{t}$ are the
coordinates of the nearest point from the fitted spiral arms to the
tracer. The Minuit package \citep{J+R+1975} was adopted to minimize
the factor $Z$. The best fitting logarithmic model of{ a} spiral
arm shows that the pitch angle of the major arms ranges from 9\deg\
to 19\deg\ (Table~\ref{tab:sac}). This is characteristic of major
arms in Sb-Sc type galaxies \citep{Kennicutt:81}.

\begin{table}  %%table4
\centering
\begin{minipage}{5cm}
    \caption{Spiral Arm Characteristics\label{tab:sac}}
\end{minipage}

 \renewcommand\baselinestretch{1.4}
\fns\tabcolsep 3mm
    \begin{tabular}{lrrr}
        \hline\noalign{\smallskip}
        Arm   &\multicolumn{1}{c}{$\beta{\rm _{ref}}$ } &\multicolumn{1}{c}{ $R{\rm _{ref}} $}& \multicolumn{1}{c}{ $\psi$ } \\
        & \multicolumn{1}{c}{($\deg$)} &\multicolumn{1}{c}{(kpc)}    & \multicolumn{1}{c}{($\deg$)}\\
(1)&\multicolumn{1}{c}{(2)}&
\multicolumn{1}{c}{(3)}&\multicolumn{1}{c}{(4)}\\
        \hline\noalign{\smallskip}
        Scutum      &   $-$3.1   &    5.9 $\pm$   0.1  &   18.7 $\pm$   0.8 \\
        Sagittarius &   $-$0.0   &    7.2 $\pm$   0.1  &   13.5 $\pm$   0.5  \\
        Local       &    2.2   &    8.5 $\pm$   0.1  &   11.5 $\pm$   0.5 \\
        Perseus     &   $-$11.8  &   10.6 $\pm$   0.1  &    9.0 $\pm$   0.1 \\
        \hline\noalign{\smallskip}
    \end{tabular}
\parbox{75mm}
{
%\begin{tablenotes}
\footnotesize \baselineskip 3.8mm
 Notes: Columns~(2)
        and (3) give the reference Galactocentric azimuth and the
        fitted radius at that azimuth{ respectively}. Column (4) is the
        spiral arm pitch angle, indicating how tightly wound the spiral
        is.}
%\end{tablenotes}
%\end{threeparttable}
\end{table}

\subsection{Fundamental Galactic Parameters}

In this section, we investigate Galactic
Parameters \Ro{ and} \To,\ solar motions and rotation
curves using both {{\it Gaia}} O star and VLBI
astrometric data. Such {a }study can examine the
consistenc{y} between {{\it Gaia}} and VLBI
techniques and the quality of current {{\it Gaia}} DR2
data. We used the Bayesian model fitting approach of
\citet{Reid:14}, based on observations of the radial velocity in the
heliocentric frame, $V_{\rm
  helio}$, and the proper motion in Galactic coordinates ($\mu_l$,
$\mu_b$).  The posteriori probability density functions (PDFs) of models were estimated with Markov
chain Monte Carlo (McMC) trials which were sampled with{ the}
Metropolis-Hastings algorithm.

In Section 3.1, the structure of the spiral arm {was}
derived from 102 masers and 635 O stars, however, to estimate
Galactic parameters \Ro\ and \To\ with kinematic models, we need to
construct a subset by removing sources with extremely large
peculiar motions. For O stars, we firstly estimated their peculiar
motions with a prior model, the Univ model from
{t}able~5 of \citet{Reid:14}{,} and subsequently
derived the standard deviation (std) of peculiar motions ($U_s$,
$V_s$, $W_s$), which are (15, 13, 10){\,}\kms,
respectively. Here $U_{s}, V_{s}$
  and $W_{s}$ are velocity components toward the GC, in the
direction of Galactic rotation, and toward the North Galactic Pole
in a Galactocentric reference frame, respectively. As $W_s$ is less
likely to be affected by the asymmetric spiral gravitational
potential, here we adopt the std of $W_s$,
10{\,}\kms as the typical value of random motions
{for} O stars.  Then O stars with peculiar motions
larger than 30{\,}\kms\ (3 times {the} random
motion) in any direction along $U_s$, $ V_s${ and}  $W_s$
components are excluded, which selects 291 O stars. With the same
criterion, we selected 95 masers. Consequently, our analysis
{was} based on the 291 O stars and 95 masers.

In \citet{Reid:14}, the proper motion and Doppler velocity weights
were given by $w(\mu)=\sqrt{\sigma^2_{\mu}+\sigma^2_{Vir}/d^2}$ and
$w$($V_{\mathrm{helio}}$) = $\sqrt{\sigma^2_{v}+\sigma^2_{Vir}}$,
where a random ({v}irial) motion of
$\sigma_{Vir}$~=~5{\,}\kms\ {was} adopted. In
this study, for the 95 maser data, we used the same weighting
strategy as \citet{Reid:14}. For the 291 O star datasets, when
calculating the observables $V_{\mathrm{helio}}$ and
($\mu_l$, $\mu_b$), we found the formal error of these observables
{is} very small, at a level of
$\sim$1{\,}\kms.  On the other hand, the std{s} of
peculiar motions ($U_s$, $V_s$, $W_s$) are (15, 13,
10){\,}\kms, which can be the typical value of
random ({v}ir{i}al) motions. Thus, in practice, we
adopted a ({v}irial) motion of
$\sqrt{(15^2+13^2)/2}=$14{\,}\kms\ in $\mu_l$ and
$V_{\mathrm{helio}}$ directions and virial dispersion of
10{\,}\kms\ in{ the} $\mu_b$ direction for O stars.
With this weighting strategy, we achieved a reduced chi-square of
$\sim$1.0, indicating such a weighting strategy is reasonable.

As to the choice of the rotation curve, by fitting Galactic
parameters and solar motions with different types of rotation curves
and comparing the posterior statistic ($\chi^2$) of these fittings,
\citet{Reid:14} concluded that the Persic96 universal rotational
curve \citep{per96} is slightly better than other models, such as
the two-order polynomial, Brand \& Blitz{ }(1993)'s (BB)
power-law \citep{bb93}, and Clemens's rotation curves \citep{cle85}.
Here we made similar comparisons of four types of rotation curves:
1st-order polynomial, 2nd-order polynomial, BB's power-law
and Persic96 universal rotational curve. We merged the 291 O stars
and 95 masers into a single dataset, which {is} used to
fit the rotation curves, and the results are listed in Table
\ref{tab:rc}. In summary, we found the same conclusion as
\citet{Reid:14} that the {u}niversal rotational curve is
better than other types of rotation curves.

Finally, we estimated parameters including \Ro, \To, rotation curve
parameters {$a2/a3$}, ({$a1$} = \To),
solar motions \U, \V, \W and average peculiar motions
$\overline{U_s}$, $\overline{V_s}$. When adopting the
{u}niversal rotation curve, we found \Ro\ = 8.35$ \pm$
0.18{\,}kpc and \To\ =~240 $\pm$ 10{\,}\kms,
which are {very} consistent with the values (\Ro\
=~8.31 $\pm$ 0.16{\,}kpc, \To\ =~241 $\pm$
8{\,}\kms) given by \citet{Reid:14} at a 1$\sigma$ level.
The angular speed $\Omega_\odot =\Theta_\odot$/$R_\odot$ of the Sun
with respect to the GC is
30.75{\,}$\pm$~0.31{\,}\kms\,kpc$^{-1}$,
which is also consistent with \citet{Reid:14}, $\Omega_\odot$ =
30.57{\,}$\pm${\,}0.43{\,}\kms\,
kpc$^{-1}$, at a 1$\sigma$ level, and consistent with the proper
motion measurement of Sgr A$^\ast$ \citep{Reid:04}, $\Omega_\odot$ =
29.45 $\pm$ 0.15{\,}\kms{\,}kpc$^{-1}$ at a
2$\sigma$ level.

\begin{figure*}  %%fig7
\centering
\includegraphics[width=12cm]{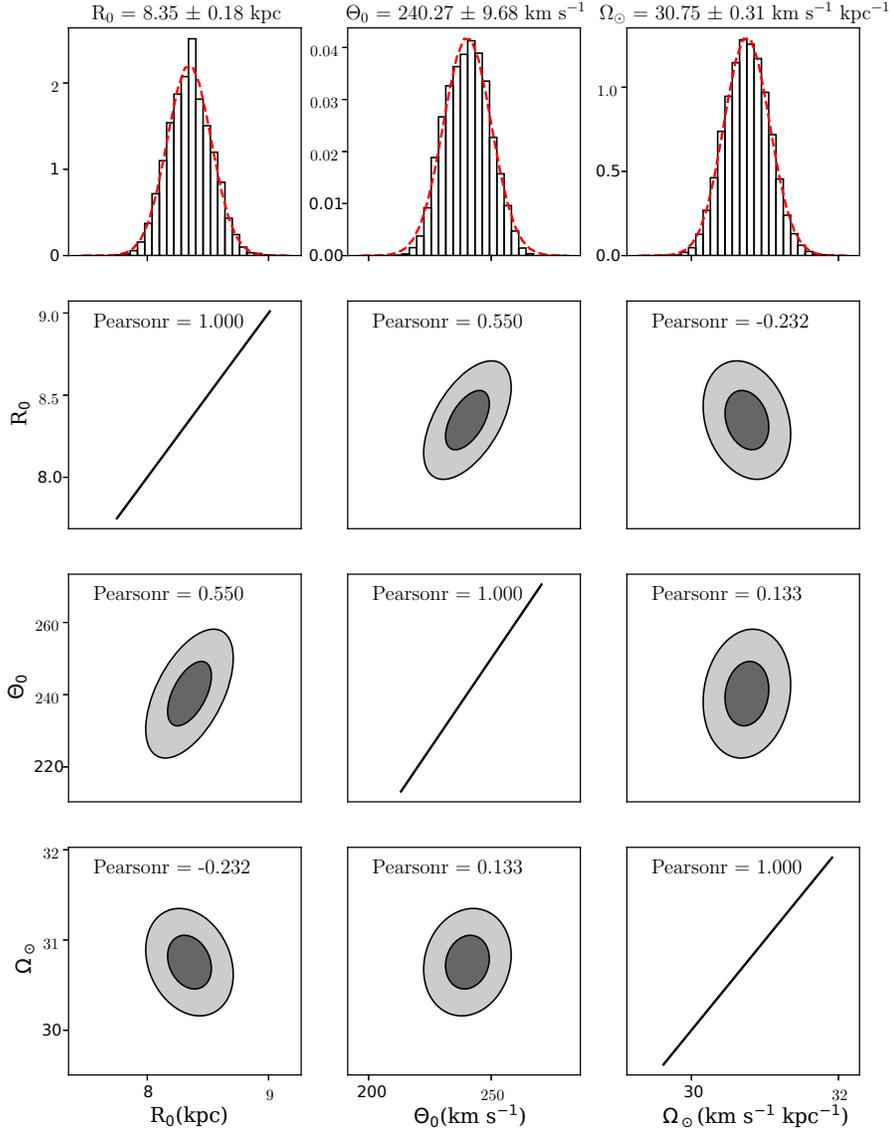}

\caption{ \baselineskip 3.8mm The marginalized (upper row panels) and joint (lower 3-row panels) posteriori {PDFs} for \Ro, \To\ and $\Omega_\odot$, estimated based on 291 O stars and 95 masers. (1) Upper panels, the red lines are Gaussian fittings of the Marginalized PDFs; (2) Lower panels, 
the deep and light grey areas denote 68\% and 95\% probabilities of Joint PDFs. Pearson correlation coefficients are  labeled in
the top of Figures.}\label{3d}
\end{figure*}

To investigate the quality of the {{\it Gaia}} DR2
data, we estimated \Ro\ and \To\ with only the 291 O star data. Here
we fixed the rotation curve parameters and solar motions, taking
into account the limited Galactic coverage of O-star data. The
fitting results {are} presented in the last column of
Table \ref{tab:rc}. The \Ro\ and \To\ values estimated with O-star
data are consistent with maser-O{ }star combined results but
with larger uncertainties. In addition, the pure O-star dataset
yields{ a} very large correlation coefficient, 0.970. In summary,
currently, maser astrometric data with better accuracy and wider
Galactic distribution are better than {{\it Gaia}} DR2
data in estimation of \Ro\ and \To. In the future, with more
measurements of maser parallaxes and better {{\it
Gaia}} datasets, we expect that the {f}undamental
Galactic parameters could be determined{ better}.

\begin{table*}  %%table5
\centering
\begin{minipage}{6cm}
\caption{Rotation Curve Fitting Results\label{tab:rc}}
\end{minipage}

\renewcommand\baselinestretch{1.45}
\fns \tabcolsep 0.6mm
  \begin{tabular}{lccccc|lllll|cccc}
\hline%\noalign{\smallskip}
      & \Ro   & \To  & r$_{R_0,\Theta_0}$ &
       $a_2$ & $a_3$ & \U           & \V & \W & \Usbar & \Vsbar
       & $N_{\rm dof}$ & $N_{\rm source}$ & $\chi^2$ \\
      & (kpc) &
      (\kms)  &                 &       &
      & \multicolumn{5}{c|}{(\kms)}% &(\kms)&(\kms)&(\kms)&(\kms)&
                &              &\\
\hline%\noalign{\smallskip}
1-Poly           &8.40$\pm$0.19&242$\pm$10&0.539 & $-$0.1$\pm$4.3 &   ---        &14.3$\pm$2.6&15.7$\pm9.0$& 8.6$\pm$0.6&5.8$\pm$2.7& 0.6$\pm$9.0 &1150&386    &1230.0 \\
2-Poly           &8.38$\pm$0.19&241$\pm$9 &0.576 &  1.7$\pm$4.0 &$-$14.8$\pm$11.0&13.9$\pm$2.7&14.6$\pm7.5$& 8.6$\pm$0.6&5.3$\pm$2.8& 0.1$\pm$7.5 &1149&386    &1225.4 \\
BB               &8.38$\pm$0.17&241$\pm$9 &0.491 &$-$0.02$\pm$0.01&   ---        &14.5$\pm$2.7&15.3$\pm8.0$& 8.6$\pm$0.6&5.8$\pm$2.7& 0.1$\pm$8.0 &1150&386    &1226.9 \\
Univ             &8.35$\pm$0.18&240$\pm$10&0.550 & 0.88$\pm$0.07&1.39$\pm$0.13 &13.3$\pm$2.6&17.0$\pm8.0$& 8.6$\pm$0.6&4.8$\pm$2.6& 2.6$\pm$8.0 &1149&386    &1211.6 \\
Univ$^\dagger$   &8.57$\pm$0.63&239$\pm$18&0.970 & 0.88         &1.39          &13.3        &17.0        & 8.6        &3.4$\pm$1.1& 3.7$\pm$1.1 & 869&291    &916.4  \\
  \hline\noalign{\smallskip}
\end{tabular}
%\begin{tablenotes}
\parbox{155mm}
{ \footnotesize Notes: $\dagger$, fit \Ro, \To\, \Usbar and
\Vsbar using 291 O stars with solar motions and rota{t}ion curve
fixed.}
%\end{tablenotes}
%\end{threeparttable}
\end{table*}

\subsection{Galactic Dynamics}

With distances, proper potions and radial velocities, one has
full 3D velocity information. In this section, we calculate 3D
velocities for these sources and demonstrate their peculiar motions
(with respect to the Galactocentric reference frame). 3D velocities
are calculated straightforwardly with linear speeds on the celestial
sphere (obtained with proper motions and distances) and radial
velocities. Subsequently, we estimate their peculiar (non-circular)
motions by subtracting {the effect of }Galactic rotation and peculiar motions of the Sun.
\begin{table*}  %%table6
\centering
\begin{minipage}{5cm}
    \caption{Mean Peculiar Motions\label{tab:mpm}}
\end{minipage}

 \renewcommand\baselinestretch{1.4}
\fns\tabcolsep 6mm
    \begin{tabular}{lcccccc}
    \hline\noalign{\smallskip}
        Source &  \multicolumn{1}{c}{$\overline{U}$}  &
        \multicolumn{1}{c}{ $\Delta U$ }& \multicolumn{1}{c}{$\overline{V}$} &   \multicolumn{1}{c}{$\Delta V$} & {$\overline{W}$}   & \multicolumn{1}{c}{$\Delta W$ }\\
        Type &
        \multicolumn{6}{c}{(\kms)}\\
        %&  \multicolumn{1}{c}{(\kms)} & \multicolumn{1}{c}{(\kms)} &   \multicolumn{1}{c}{(\kms)} & \multicolumn{1}{c}{(\kms)} & \multicolumn{1}{c}{(\kms)}\\
        \hline\noalign{\smallskip}
        Masers &    4.9 &   11.9 &    1.6 &   10.0 &    2.0 &    8.2 \\
        O stars&    8.1 &   19.6 &    8.6 &   16.8 &    3.5 &   10.9   \\

        \hline\noalign{\smallskip}
    \end{tabular}
%\begin{tablenotes}
\parbox{105mm}
{ \footnotesize
        %\item[]
        Notes: $\overline{U_s}$,
                  $\overline{V_s}$ and $\overline{W_s}$ are velocity
                  components toward the GC in the direction of
                  {G}alactic rotation and toward the {N}orth {G}alactic
                  {P}ole
                  average peculiar motions, respectively. The mean
                  peculiar motions of both the O stars and the masers
                  are small, indicating the motions are
                  random. However, O stars have much larger {std} ($\Delta U$, $\Delta V$) than that of
                  masers.}
%\end{tablenotes}
%\end{threeparttable}
\end{table*}

\begin{figure*}  %%fig8
\centering
\includegraphics[width=8cm]{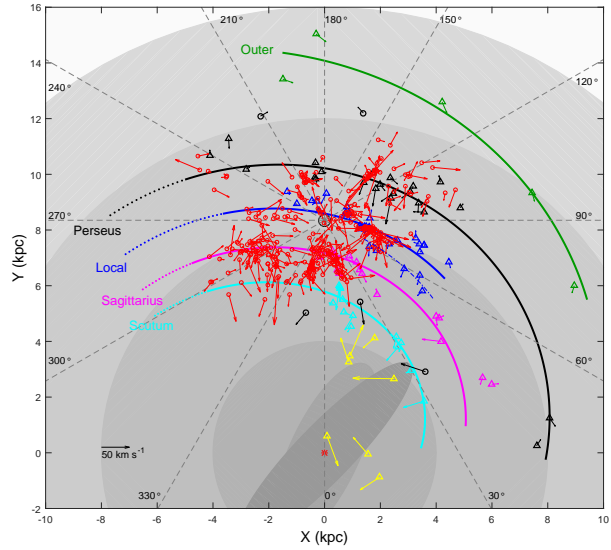}

\caption{ \baselineskip 3.8mm Peculiar motion vectors of O stars
({\it open circles}) and masers ({\it open triangles}). A motion scale of 50{\,}\kms\ is indicated in the
bottom left
  corner of the panel. The background is the same as {in} figure~1 of
  \citet{Reid:14}.  The Galaxy is viewed from the {N}orth
    Galactic {P}ole, it rotates clockwise and the Sun is at ($X=0.0${\,}kpc,
    $Y=8.35${\,}kpc).}
\label{3d}
\end{figure*}

We estimated peculiar motions of O stars following \cite{Reid:09b},
using updated Galactic parameters of 240{\,}\kms\ for the
Galactic rotation speed, $\Theta_{0}$, at a distance of 8.35\,kpc to
the GC, $R_{0}$, and solar motion parameters of $U_{\odot} =
13.3${\,}\kms, $V_{\odot} =
17.0${\,}\kms and $W_{\odot} =
8.6${\,}\kms\ from this work (``{u}niversal''
rotation curve model). Because velocities for bright stars are not
available in {{\it Gaia}} DR2, the velocities of O
stars are taken from \cite{reed03}.

Since sources near the GC may have large non-circular motions, due
to the great gravitational potential of the Galactic bar, we removed
the sources within a radius of 4\,kpc from the GC. In addition, we eliminated those with peculiar motions in the
direction perpendicular to{ the} Galactic plane, larger
than 40{\,}\kms\ and{ with} a 3D velocity of more
than 90{\,}\kms, resulting{ in} a total of 318 O
stars and 94 masers. The results of the peculiar motions of the sources are
listed in Table~\ref{tab:sac}, while the peculiar motions are shown
in Figure~\ref{3d}. %0000000000000

One can see that the average velocities of O stars are larger than
{those} of masers, especially in the direction of
Galactic rotation. O stars rotate faster, $>$8{\,}\kms,
than the rotational speed, while the maser's movement approximates
the rotation speed. On the other hand, {i}t is noted that
although their peculiar motions are generally random, toward the GC
and the direction of Galactic rotation, the O stars have much larger
std than that of the masers. Here the gravitational potential of the
Galactic bar is small, so the peculiar motions are intrinsic,
suggesting that O stars and masers may be located in different
physical environments. Because these masers are associated with
{HMSFRs}, when the high-mass
stars should be bound to their native molecular clouds tightly,
some of the O stars have perhaps picked up peculiar
motions from the dispersal of gas/dust from their birth clouds to
move out there.

We further investigate the number of O stars associated with their
natal molecular clouds. Based on the archived CO data\footnote{{\it
http://www.radioast.nsdc.cn/english/index.php}} with a typical
{root mean square (}rms{)} noise level of about 0.3\,K , we
found 207 O stars have CO observations and 103 of them have
associated $^{13}$CO emission within 10$\arcmin$, indicating that
about 50\% (1 $-$ 103/207){ of} O stars may have left their
natal molecular clouds already. In addition, we found a large
average deviation of $\sim$16{\,}\kms\ between
remaining O stars and the CO molecular clouds. Since CO emission is
widespread in the Galactic plane, in order to avoid the ambiguity
caused by multi{ple }peaks in CO spectra, we only use the
CO data with {a }single peak. Such a deviation indeed suggests
that the O stars and masers live in different physical
environments.

\section{Prospects}

\subsection{Limitations of Present Facilities}
Although the VLBI technique can achieve an angular
accuracy of a few $\upmu$as, it is not easy to fully reconstruct the
Galactic spiral arms {because of} many
limit{ing} factors. At present, the astrometric errors are large
due to the sensitivity of current equipments, atmospheric effects
and lack of stations in{ the} southern hemisphere. The poor
sensitivity of telescopes, such as the VLBA, results in limitation
of available sources. Because phase-referenced observations 
involve two angularly-close sources, i.e., a phase calibrator
({q}uasar) and a target source (maser, pulsar, etc), the
switching time must be very short, especially at high frequencies,
usually a minute or less to achieve phase connection across multiple
scans. Therefore, only the sources with strong intensities can be used for parallax measurements with the VLBA. In practice,
12{\,}GHz methanol and 22{\,}GHz water masers
that are roughly stronger than 5{\,}Jy are useful. In
this case, in total only about 300 of both kinds of masers are
available whereas the{ir} total number is over
3000. The switching time is a little bit longer for 6.7{\,}GHz methanol masers, so the flux density threshold is
$\sim$2{\,}Jy. In this case, approximately 400 6.7{\,}GHz methanol masers are useful (also in consideration of
compactness of sources), while the expected total number of such
masers is over 3000. Therefore, finishing these scientific aims
requires improved sensitivity to the present equipments.

At present, atmospheric effects dominate the astrometric errors.
{The i}n-beam calibration method could greatly
remove atmospheric effects. However, low sensitivity and
small field of view of present instruments make it hard to find
in-beam calibrators. Due to atmospheric effects, systematic errors
are proportional to the separation angle between targets and their
calibrators. For example, there are several calibrators for the
{HMSFR} W3OH. Among them,
one calibrator has a separation of less than 1 degree, which leads
to an accuracy of about 15{\,}$\upmu$as, while another
calibrator doubles in angle, resulting in the uncertainty almost
doubl{ing} \citep{xrzm06}. Therefore, to obtain
high{ly} accurate astrometry, it is necessary to use
high{ly} sensitive telescopes and many more calibrators close to
targets.

In addition, there are {relatively} few stations in
the southern hemisphere. In order to improve the
{$u-v$} coverage of interferometry and image
quality, both southern and northern hemisphere telescopes are
needed. At present, only{ the} Australia Long Baseline Array
(LBA) is{ being} test{ed} to measure the parallaxes
toward some star forming regions in the southern hemisphere.
However, with its limited numbers of antennas and short baselines,
it may not have competitive advantages {in} this
project.

\subsection{High{ly} Accurate Astrometry with the SKA}

The {Square Kilometre Array (}SKA{)} will change this
situation revolutionarily. Although exact types and numbers of
antennas have not been determined yet, one square kilometer of
collecting area is the final aim. SKA-VLBI sensitivity would be
expected to achieve a sensitivity of $\upmu$Jy flux, roughly two
orders of magnitude better than now. This should significantly
increase the ability to detect weak sources. Therefore, it can find
many more targets and calibrators with accurate positions because
sources with high signal-to-noise ratios can greatly improve their
position accuracies. On the other hand, the antenna{s}
{composing} the SKA {are} not large, about
15-m in diameter, so it has{ a} relative{ly} large field of
view.

The superior sensitivity and large field of view will ensure detect{ing} more targets and calibrators within the same
beam. For baselines over 1000{\,}km, the systematic
errors are dominated by delays introduced by Earth's
atmosphere and ionosphere. For high{ly} accurate astrometry, it
is crucial to remove the residual tropospheric and ionospheric
effects, in particular for low-frequency observations. In general,
there are two ways: 1) measuring the tropospheric delay above each
antenna during observations. By observing large
numbers of extragalactic sources spread over the sky to measure
broad-band delays \citep{Reid:09a}{,} {t}he
ionospheric delays can be partially removed by applying a global
ionospheric model derived from GPS measurements; 2) using calibrators with a small
angular separation from targets. Sources as weak as a few $\upmu$Jy
should be useful for the SKA-VLBI observations. Statistically, the
weaker the calibrator{ is}, the higher the chance
{it can} be found near the target. Using adjacent
(in-beam calibration is optimal) and multiple calibrators helps
reduce systematic errors owing to time variation in{ the}
atmosphere and calibrator structure. In this way, a good imaging
quality would be produced and the astrometric accuracy will be
roughly equal to the resolution divided by the dynamic range of the
image. Therefore, together with current VLBI arrays, one
can obtain an accuracy of a few $\upmu$Jy for positions and
$\sim$1{\,}$\upmu$Jy for parallaxes at high
frequencies ($>$5{\,}GHz), which ensures parallax and
proper motion measurements throughout the whole Milky Way.

\section{Summary}

There is {neither} general agreement on the number of
arms nor on their locations and orientations in early models of
spiral structure because typical uncertainties in distances are
comparable to the spacing between arms. With the most accurate
astrometric parameters of the youngest O-type stars and
parallax-measured masers, for the first time, the spiral structure
in all four quadrants {is} delineated clearly. The
revealed Galactic spiral patterns make a clear sketch of nearby
spiral arms. In addition, the best values of \Ro\ and \To\ were
estimated. However, the progress {on} VLBI
astrometry is largely limited by low sensitivity of present
facilities and large residual atmospheric effects. The superior
sensitivity and large field of view of the SKA will allow us to map
objects {with} unprecedented accuracy throughout the
entire Galaxy.

\normalem
\begin{acknowledgements}
This work was sponsored by the MOST (Grant
No.~2017YFA0402701), the NSFC (Grant Nos.~11873019,
11673066 and 11503033), the CAS (Grant
No.~QYZDJ-SSW-SLH047), the Youth
Innovation Promotion Association of CAS and is also supported by the Key
Laboratory for Radio Astronomy, CAS and the Open
Project Program of the Key Laboratory of FAST, NAOC, CAS. This work has made use of data from the ESA mission {\it
  Gaia} ({\it https://www.cosmos.esa.int/Gaia}), processed by the {\it
  Gaia} DPAC,
{\it https://www.cosmos.esa.int/web/Gaia/dpac/consortium}. Funding
for the DPAC has been provided by national institutions, in
particular the institutions participating in the {\it Gaia}
Multilateral Agreement.

\end{acknowledgements}

%\bibliographystyle{/home/lixh/raa/bibstyle/raa_apj_nolink}
%\bibliography{bibtex}
\end{document}